\documentclass[aps, pre,reprint,nofootinbib,citeautoscript,floatfix]{revtex4-1}
\pdfoutput=1

\usepackage{mathtools}
\usepackage{amsmath, amssymb, amsthm, amsbsy, bbm}

\usepackage[percent]{overpic}
\usepackage[dvipsnames]{xcolor}
\usepackage[utf8]{inputenc}
\usepackage{enumitem}
\usepackage{wasysym} 
\usepackage{hyperref}
\usepackage{graphicx}

\usepackage{subfig}

\usepackage{tikz}

\newcommand{\customyinyang}[1][1]{%
    \begin{tikzpicture}[scale=#1*0.07]
      \draw[line width = #1*0.12mm,transform canvas={yshift=0.02cm}] (0,0) circle (1cm);
      \path[fill=black,transform canvas={yshift=0.02cm}] (90:1cm) arc (90:-90:0.5cm)
                        (0,0)    arc (90:270:0.5cm)
                        (0,-1cm) arc (-90:-270:1cm);
    \end{tikzpicture}}

\usepackage[normalem]{ulem}
\usepackage{etoolbox}
\usepackage{cancel}
\usepackage{todonotes}
\usepackage{algpseudocode}

\newtheorem{theorem}{Theorem}

\usepackage[many]{tcolorbox}    

\tcbset{
    colback = blue!5!white,
    colframe=blue!15!black,
    before skip = 0.2cm,    
    after skip = 0.5cm      
} 

\newtcolorbox{example}{
    boxrule = 0.5pt,
    toprule = 0.5pt, 
    enhanced,
    fonttitle=\bfseries,
    breakable
}

\begin{document}

\title{Inference, interference and invariance: \\ 
How the Quantum Fourier Transform can help to learn from data}

\author{David Wakeham}
\author{Maria Schuld}
\affiliation{Xanadu, Toronto, ON, M5G 2C8, Canada}

\begin{abstract}
How can we take inspiration from a typical quantum algorithm to design heuristics for machine learning? A common blueprint,
used from Deutsch-Josza to Shor's algorithm, is to place labeled information in superposition via an oracle, 
interfere in Fourier space, and measure.
In this paper, we want to understand how this interference strategy can be used for \textit{inference}, i.e. to generalize from finite data samples to a ground truth. Our investigative framework is built around the  
Hidden Subgroup Problem (HSP), which we transform into a learning task by replacing the oracle with classical training data. 
The standard quantum algorithm for solving the HSP uses the Quantum Fourier Transform to expose an invariant subspace, i.e., a subset of Hilbert space in which the hidden symmetry is manifest. Based on this insight, we propose an inference principle that ``compares'' the data to this invariant subspace, and suggest a concrete implementation via overlaps of quantum states. We hope that this leads to well-motivated quantum heuristics that can leverage symmetries for machine learning applications.

\end{abstract}

\maketitle

\section{Introduction}

The Hidden Subgroup Problem (HSP) \cite{Hallgren2003, Ettinger_2004}
is the task of discovering a subgroup from information about the way it partitions the parent group.
While abstract, it neatly generalizes many problems solved by quantum algorithms, from Deutsch-Jozsa \cite{DJ} to Simon's
problem \cite{Simon} to Shor's algorithms for period-finding and discrete logarithms \cite{shor}.
The standard quantum routine for the HSP \cite{Childs_2010} has a common and embarrassingly
simple blueprint: label all inputs, uniformly superpose, apply the
Fourier transform, and measure. Formally, this samples from a rather abstract group-theoretic object called the \textit{annihilator} of the hidden subgroup, and together with some classical post-processing, allows us to find the subgroup exponentially faster than any classical algorithm.

Given the simplicity and power of this strategy, it is natural to wonder if the underlying mechanism can be applied to practical learning tasks. In broad terms, we ask the following:
\begin{quote}
    \textit{How can a quantum computer's unique access to information in Fourier space help us learn from data?}
\end{quote}

In this paper, we initiate a research program which aims to answer this question. Our initial playground will be a learning variant of the HSP where, instead of using an oracle to prepare a dense state which encodes the full structure of the task, we only have access to few, randomly sampled computational basis states from this superposition.
With this framework, we can make the question above more precise:
\begin{quote}
    \textit{Can access to the annihilator of a subgroup help us decide from which ``true'' oracle the training data was sampled?}
\end{quote}

\begin{figure}[t]
  \centering
  \includegraphics[width=0.8\columnwidth]{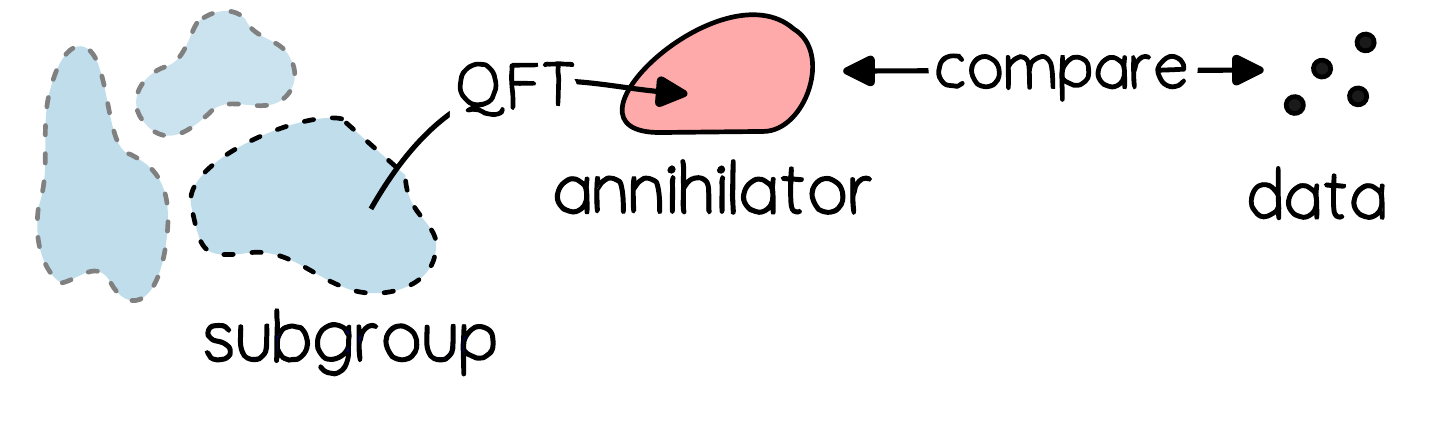}
  \caption{Which hidden subgroup gave rise to the data? We propose to compare the data with the annihilator of a given subgroup. The annihilator is computed via a group Quantum Fourier Transform executed by a quantum computer. }
  \label{fig:overview}
\end{figure}

Inspired by the algorithm which solves the original HSP, we develop an inference principle---i.e., a recipe that tells us how likely it is that a given hidden subgroup has produced the data---based on the subgroup's annihilator (Fig.~\ref{fig:overview}). The essential idea is to compare the subspace of states spanned by the annihilator, which is invariant with respect to the hidden subgroup, with a subspace of quantum states representing the data. The quantum computer enables this comparison by providing fast implementation of the Quantum Fourier Transform (QFT) and hence fast access to the crucial annihilator subspace. 

\vspace{0pt}
\vspace{0pt}

Our work relates to several 
broader themes in quantum machine learning. One such theme is \textit{geometric quantum machine learning} \cite{Meyer_2023, GIQML, bronstein2021geometric}, which focuses on how to make variational circuits invariant with respect to certain group symmetries. In contrast, we do not seek a symmetry-aware circuit design, but try to understand how symmetries can become the basis of inference itself. We also focus on the realm of fault-tolerant quantum algorithms rather than variational circuits. Another related theme is \textit{quantum learning theory} (i.e., \cite{arunachalam2017survey, caro2023classical, dunjko2017exponential}) which typically uses Fourier sampling in the oracular setting to prove quantum advantages for learning. While we do occasionally borrow from these techniques, we replace oracles (which provide access to an entire data distribution) with the more realistic case of finite classical training data. Our question is not whether our problem admits a quantum speedup, but how the information in Fourier space can be used by a learner.\footnote{Of course, mathematical proofs can sometimes help to build intuition for algorithmic design, but we did not find this to be the case with the typical results in quantum learning theory.}

This unusual choice---to neither develop end-to-end algorithms nor prove quantum advantages--- is a conscious decision. Ultimately, we are interested in ``real world" data processed on full-scale quantum hardware. The structures that give rise to speedups are only approximately and messily realized in such data (and we might not even \textit{know} if they are present in a given task). Quantum machine learning research therefore needs to find ways of developing carefully motivated heuristics, which we expect will look very different from the limited set of quantum algorithms whose performance we can rigorously guarantee. But to identify promising heuristics, we need to understand the mechanisms of quantum learning qualitatively: \textit{how} is 
at least as important as
\textit{how much better}. Having said that, we select the HSP as our starting point precisely \textit{because} it shows the most promise for exponential speedups, both in the non-black box \cite{shor} and black box settings \cite{aaronson2009bqp, aaronson2014forrelation}. Fourier sampling is also the recurring recipe behind most speedups for quantum learning proven so far \cite{caro2023classical, dunjko2017exponential, Liu_2021}. We therefore hope that heuristics derived in this playground are good candidates to exploit the unique contribution of quantum information processing to practical machine learning.

\subsection{Outline of the paper}
\begin{figure}[t]
  \centering
  \includegraphics[width=0.9\columnwidth]{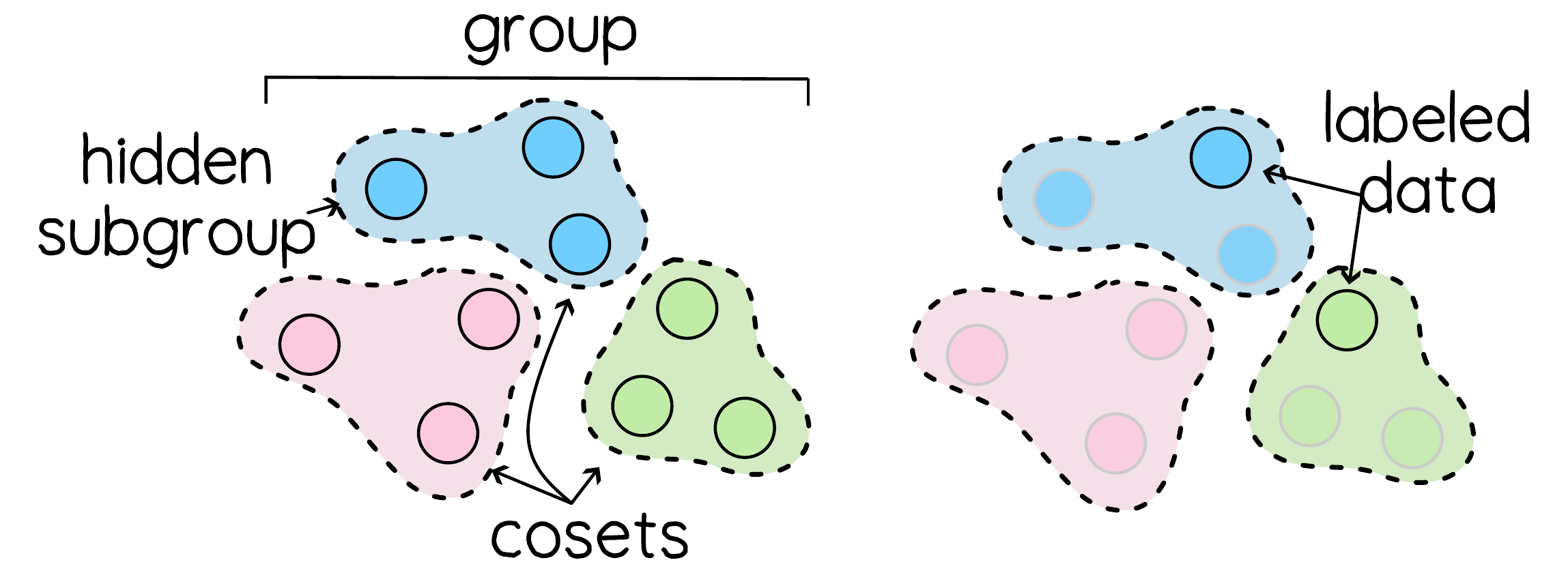}
  \caption{\emph{Left}. In the Hidden Subgroup Problem, a
    labeling function partitions a group into copies of a hidden
    subgroup called \textit{cosets}. \emph{Right}. In the learning variant,
    we are given few labeled samples from cosets.}
  \label{fig:hsp-illustration}
\end{figure}

Since the content of this investigation is of rather technical nature, we want to first summarise the results, and where they can be found, in more detail.
Our starting point is to convert the HSP into a learning problem.
We replace the quantum oracle $U_f$, which implements a classical function $f$ that carries information on how the subgroup partitions its parent group, with a set of samples labeled by $f$ (Fig.~\ref{fig:hsp-illustration}).
We will see that learning $f$ corresponds to the original task of guessing the hidden subgroup (\S\ref{sec:hsp-clustering}).

\vspace{0pt}
\vspace{0pt}

For realistically-sized training sets, the standard quantum algorithm no longer yields the hidden subgroup, since without the complete oracle, amplitude leaks out of the annihilator and into other states. What we need instead is a learning algorithm that generalises from the data samples to \textit{infer} the right subgroup. Encouragingly, we show that in the framework of PAC learning, the problem has efficient sample complexity, which means that few data samples contain enough information to solve it  (\S\ref{sec:sample}). This little exercise tells us that in principle, learning from a finite dataset is possible. Note that efficient sample complexity does \textit{not} tell us how to construct a learning algorithm.\footnote{In fact, it does not guarantee an efficient quantum or classical learning algorithm even \textit{exists}, which is an open question of academic interest. As mentioned before, we are seeking heuristics and approximate solutions here and will not worry any further about the possibility that the optimal solution is unobtainable.}  

How can we access the information contained in the training set using the unique properties of quantum computers? Or, in more technical terms: \textit{How can access to annihilators from different candidate subgroups help to find the correct subgroup from data?} In Section (\S\ref{sec:reveal}) we present the main results of the paper: we identify a promising building block of a heuristic quantum machine learning algorithm for the learning variant of the HSP. The building block is an inference principle that compares data to an invariant subspace spanned by an annihilator: a subgroup is considered more likely to produce the data if the distance to its annihilator's invariant subspace is smaller. For the perfect oracle from which the data was sampled, this distance is provably zero. 
We show how to practically implement this principle by computing a surrogate measure we call the \textit{data-annihilator overlap}, which can be used as a cost function whose evaluation is facilitated by quantum computers.

Although finding hidden 
subgroups may seem like a specialized affair---most relevant to breaking cryptographic protocols or discovering physical laws---we argue that learning the HSP may be useful for the generic task of understanding how variation splits into task-relevant factors and those that are merely nuisances
(\S\ref{sec:dao-deep-learning}).  

As mentioned above, this paper is only a first step of a larger research program. The conclusion (\S\ref{sec:conclusion}) will therefore summarize key take-aways to go beyond the artificial setting of hidden subgroup problems in future.

\section{Learning and the HSP}
\label{sec:hsp-clustering}

We will start with a brief overview of group theory and the HSP, then introduce the learning variant of the problem which we recast as a classification task. Finally, we outline the standard quantum algorithm for solving the HSP, and see why it fails for the learning variant of the problem.
The key takeaway is that the Quantum Fourier Transform reveals a set of objects called the \textit{annihilator} of the hidden subgroup. The standard quantum algorithm samples from the annihilator, but for a finite data set this is no longer the case.

\subsection{The Hidden Subgroup Problem}

The HSP involves a mathematical structure called a \emph{group} $G$, consisting of a set of elements (representing
symmetry transformations) and a binary composition law (representing the result of applying one transformation after the other). We will always assume that $G$ is \emph{abelian}, so
the order of composition is irrelevant. In this case, we denote
composition by $+$, and have
\[
  g + g' = g' + g
\]
for any $g, g' \in G$.
A \emph{subgroup} $H \leq G$ is a set of transformations which
form a group in their own right. 
A \emph{coset} is a shifted copy of the subgroup, i.e. a set of the form
  \begin{equation}
  r + H = \{r + h : h \in H\}.\label{eq:11}
\end{equation}
The set of (distinct) cosets of $H$ in $G$ is denoted $G/H$.
For a more formal introduction to group theory, see Appendix \ref{sec:group-theory} or abstract algebra textbooks such as \cite{Dummit1999AbstractA, lang02, algebra}.

\begin{example}
We take ``clock arithmetic" as our running example, with group $G = \mathbb{Z}_{12} = [12] := \{0, 1,
2,\ldots, 11\}$ and operation addition modulo $12$.
We choose the subgroup $H = 2\mathbb{Z}_{12} = \{0, 2,
\ldots, 10\}$.

\begin{center}
\includegraphics[scale=0.27]{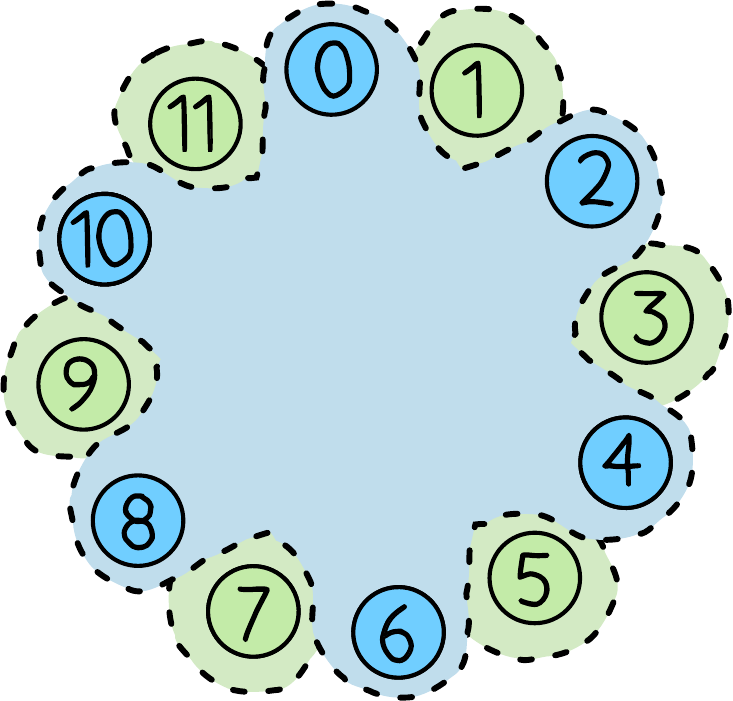}
\end{center}

The cosets are the even integers $H$ (blue) and odd integers $1 + H =
\{1, 3, \ldots, 11\}$ (green).

\end{example}

It can be shown that cosets do not overlap and cover $G$, i.e. they
``tile'' the full space $G$ (Fig.~\ref{fig:hsp-illustration},
left).
The HSP is the problem of identifying the subgroup
from a function or ``oracle'' that reveals the tiling structure (i.e., the colours in the preceding example). More formally, for a subgroup $H$ with
cosets $G/H$, and an oracle $f: G \to S$ which takes
distinct, constant values on different cosets,
  \begin{equation}
  f(g) = f(g') \quad \Longleftrightarrow \quad g - g' \in H,\label{eq:label}
\end{equation}
our task is to find the subgroup $H$.

\begin{example}
When $G = \mathbb{Z}_{12}$, $f$ maps integers to the label set
$S =\{\textcolor{cyan}{\bullet}, \textcolor{lime}{\bullet}\}$,
with $f(g) = \textcolor{lime}{\bullet}, \textcolor{cyan}{\bullet}$ 
for odds, evens.
\end{example}

\subsection{Learning variant of the HSP}

While quantum algorithms usually assume full oracle access to $f$, machine learning typically assumes access to small set of samples of $f$. The natural learning variant of the HSP therefore replaces
$f$ by data:
\begin{equation}
  \label{eq:2}
  \mathcal{T} = \{(g, f(g)): g \in X\},
\end{equation}
where $X \subseteq G$ is a set of $N = |X|$ observed points.
The supervised learning task is to guess the label $f(g')$ for an unseen data point $g'$.
In Fig.~\ref{fig:hsp-illustration}, labeling unseen data means
assigning a color, or equivalently, deciding which class it belongs
to.
Thus, we can view the HSP as a highly structured multi-class classification task.

\begin{example}
  Suppose our data is $\mathcal{T} = \{(0,
  \textcolor{cyan}{\bullet}), (2, \textcolor{cyan}{\bullet}), (3,
  \textcolor{lime}{\bullet})\}$, or
  
\begin{center}
\includegraphics[scale=0.27]{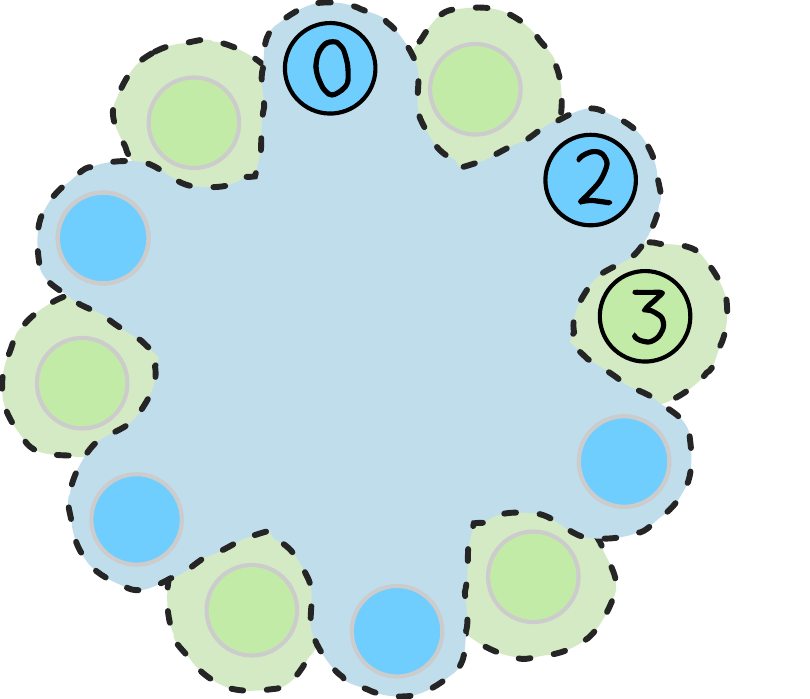}
\end{center}

Since the hidden subgroup must include $2-0=2$, but not $3-2=1$, we must have $H = 2\mathbb{Z}_{12}$.
\end{example}

While the interpretation of the HSP as a multi-class classification task is straight forward, we will occasionally make use of a different formulation which casts the problem as a binary classification task derived from clustering. The reason is that strictly speaking, only the structure of the classes themselves, i.e. the cosets, matters. This can be understood as unsupervised \emph{clustering}, or grouping inputs without knowledge about what the groups represent.
Instead of trying to guess the label of an element $f(g) \in S$, we can therefore consider the binary classification task of trying to guess if two elements are in the same cluster or not, $f(g) = f(g')$ or $f(g) \neq f(g')$, or equivalently $g \sim g'$ or $g \not\sim g'$.

Formally, for a subgroup $H$, we define the binary function $R_H: G
\times G \to \{\texttt{0}, \texttt{1}\}$ by
\begin{equation}
  \label{eq:R_H}
  R_H(g, g') =\mathbb{I}(g -g' \in H),
\end{equation}
where $\mathbb{I}$ is the indicator function and
$\texttt{0}$ and $\texttt{1}$ are Booleans.
Our labeled training data $\mathcal{T} = \{(g, f(g))\}_{g\in X}$ can be
transformed into binary training examples:
\begin{equation}
  \mathcal{T}_\text{bin} = \{(g, g', \mathbb{I}(f(g) = f(g'))) : g, g' \in X\}\label{eq:9}.
\end{equation}

\begin{example}
The labeled training set for $G = \mathbb{Z}_{12}$ becomes
  \[
    \mathcal{T}_\text{bin} = \{(0, 2, \texttt{1}), (0, 3, \texttt{0}), (2, 3, \texttt{0})\}
  \]
in the binary formulation.
A point is always in the same cluster as itself, so we omit trivial self-relations.
We illustrate the binary data below:
\begin{center}
\includegraphics[scale=0.27]{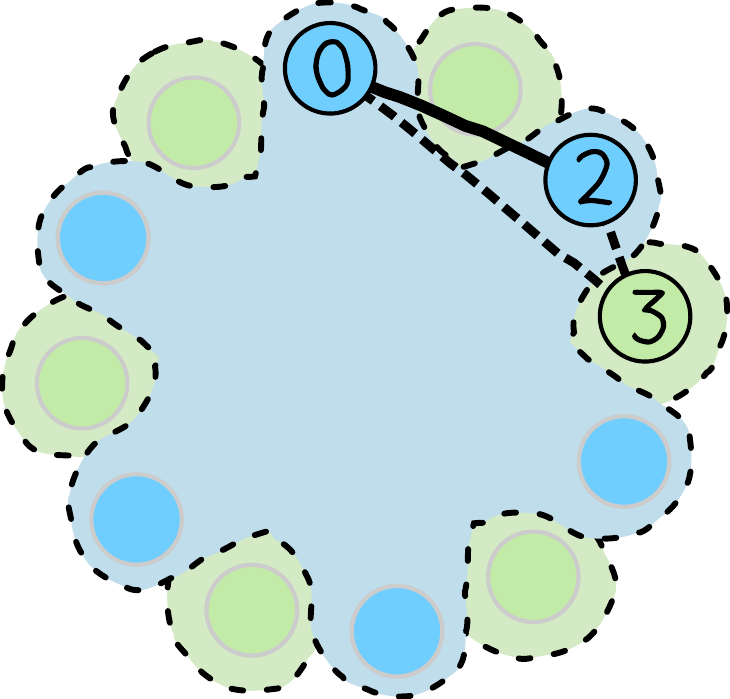}
\end{center}
A solid line means elements are in the same cluster, and a dotted line
means different clusters. 
\end{example}

\subsection{The HSP algorithm}
\label{sec:hsp-algorithm-1}

We now give an overview of the standard quantum algorithm solving the HSP, following \cite{Childs_2010}.
The algorithm will act on a tensor product
$\mathcal{H}_G\otimes \mathcal{H}_S$ of Hilbert spaces associated with $G$ and $S$.
The oracle $f$ is encoded as a unitary $U_f$ with action
  \begin{equation}
  U_f |g, s_0\rangle = |g, f(g)\rangle\label{eq:oracle}
\end{equation}
for an initial state $|s_0\rangle \in \mathcal{H}_S$.

The goal of the standard quantum algorithm is to use the oracle to learn $H$.
The algorithm first collects all labeled data by applying
the oracle to a uniform superposition. It then applies a
Quantum Fourier Transform, exposing the annihilator mentioned above, and measures.
Let us go through these steps in detail and see why they reveal the subgroup $H$.

\subsubsection{Preparing a coset state}

First, we build a uniform superposition $\sum_{g \in G}|g\rangle$ of
basis states $|g\rangle$,  corresponding to elements $g \in
G$, and apply the oracle $U_f$. This entangles every group element with its function value,
$|\Psi_f\rangle \propto \sum_{g \in G}|g, f(g)\rangle$.
If we factorize the equation according to the value of $f(g)$, we obtain
\begin{equation}
    \label{eq:r+H}
  |\Psi_f\rangle
  =\frac{1}{\sqrt{|R|}}\sum_{r\in R}|r+H, f(r)\rangle
\end{equation}
for a set $R$ of representatives from each coset.
Since each coset has size $|H|$, there are $|R| = |G|/|H|$ cosets. Note that here and going forward we use the shorthand $|X\rangle$ for a uniform superposition of kets $|x\rangle$, $x \in
X$.

Discarding the $S$ register (either by measuring or tracing out) gives us a \emph{coset state}
\begin{equation}
    |r+ H\rangle = \frac{1}{\sqrt{|H|}} \sum_{h \in H} |r + h\rangle \label{eq:coset}
\end{equation}
for uniformly random $r \in R$. 

\begin{example} 
For $G = \mathbb{Z}_{12}$ and $H = 2\mathbb{Z}_{12}$, we
can choose a set of representatives $R = \{0, 1\}$, so that
\[
  |\Psi_f\rangle = \frac{1}{\sqrt{2}}|H, \textcolor{cyan}{\bullet}\rangle +
  \frac{1}{\sqrt{2}}|1 + H, \textcolor{lime}{\bullet}\rangle.
\]
Ignoring the second register, we have an even chance of observing
$|H\rangle$ or $|1+H\rangle$.
\end{example}

\subsubsection{The group QFT}

The next step is to apply the Quantum Fourier transform (QFT)
$\mathcal{F}$. 
This involves a classical object called the \emph{dual group} $\hat{G}$,
consisting of all multiplicative functions $\chi: G \to
\mathbb{U}(1)$, satisfying
\begin{equation}
  \label{eq:35}
  \chi(g + g') = \chi(g)\chi(g').
\end{equation}
These multiplicative functions are called \emph{characters}.
Specific subsets of characters, the annihilators, will play a starring role in this paper.

For now, we note that under pointwise multiplication
$(\chi\chi')(g) = \chi(g)\chi'(g)$, the characters also form a group. As we show in Appendix \ref{sec:QFT}, the dual group is the same size as the original
group, $|\hat{G}| = |G|$, so we can index elements $\chi_y \in
\hat{G}$ using elements $y\in G$.\footnote{To avoid confusion, we use letters from the end of the alphabet when group elements are used as labels.}
We can also form an alternative basis for the  Hilbert space $\mathcal{H}_G$, with basis states $|\hat{y}\rangle$ defined by
\begin{equation}
    \langle \hat{y} | g\rangle = \frac{1}{\sqrt{|G|}}\chi_y(g). \label{eq:overlap}
\end{equation}
The QFT, denoted $\mathcal{F}$, is simply a unitary transformation implementing the change from the $|g\rangle$ basis to the $|\hat{y}\rangle$ basis:
\begin{equation}
    \mathcal{F} |y\rangle = |\hat{y}\rangle = \frac{1}{\sqrt{|G|}}\sum_{g\in G}\overline{\chi_y(g)}|g\rangle,
\end{equation}
where we inserted a resolution of identity in the $|g\rangle$ basis for the last equality.

\begin{example}
For $G = \mathbb{Z}_{12}$, the characters are exponential functions 
 $\chi_y(x) = e^{2\pi i xy/12}$, and the QFT $\mathcal{F}$ is
\begin{align*}
  \mathcal{F} & = \sum_{y\in \mathbb{Z}_{12}} |\hat{y}\rangle \langle y| = \frac{1}{\sqrt{12}}\sum_{y,x \in \mathbb{Z}_{12}} e^{-2\pi i xy/12}|x\rangle \langle y|.
\end{align*}
Recall that $|\hat{y}\rangle$ has coefficients $\overline{\chi_y(x)}$, not
$\chi_y(x)$. We can think of characters as ``ticking'' off phase at a
rate of $2\pi y/12$ radians per ``hour'':
\vspace{0pt}
\begin{center}
\includegraphics[scale=0.27]{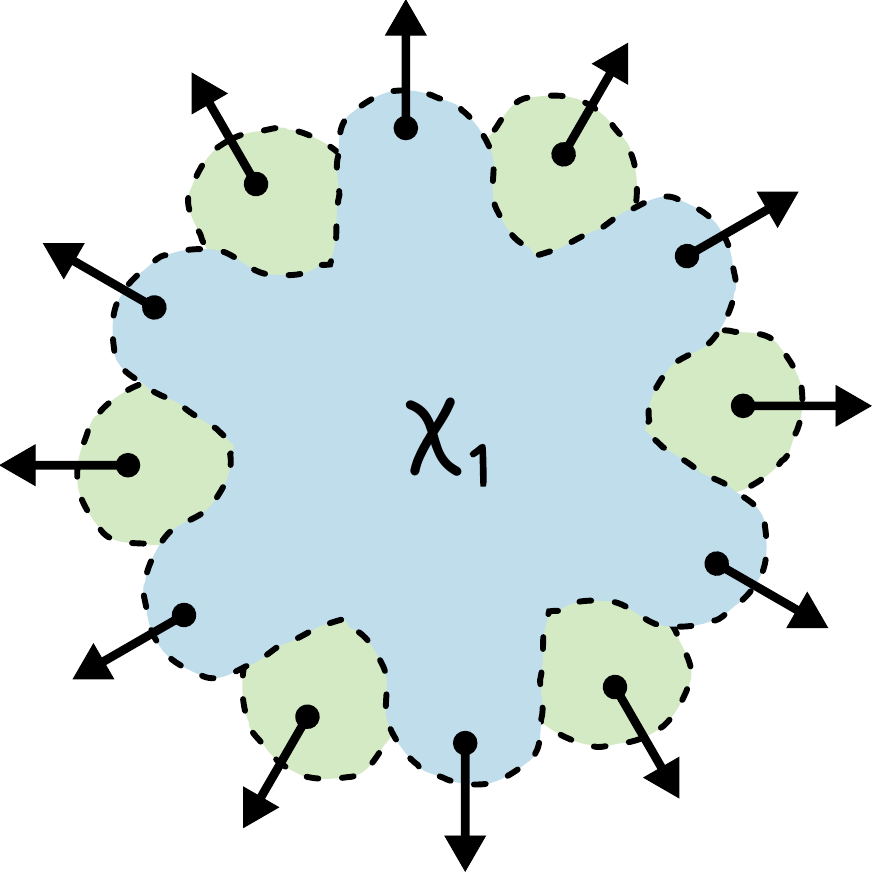}
\end{center}
\vspace{0pt}
Here, phase ticks clockwise and from the vertical, as is
conventional for clocks.
\end{example}

\subsubsection{Fourier sampling}

Applying the QFT and measuring is equivalent to measuring in the
Fourier basis $|\hat{y}\rangle$, also called \emph{Fourier sampling}.
Our algorithm starts by preparing a coset state, and terminates by
Fourier sampling.
Using (\ref{eq:coset}), (\ref{eq:overlap}), and the multiplicativity
of characters, the probability of observing $\hat{y}$ (or sampling character $\chi_y$) for $|r +
H\rangle$ is
\begin{align}
\hat{p}(y) =  |\langle \hat{y} | r+ H\rangle|^2 & = \frac{1}{|H|}\left|\sum_{h\in H}
                                      \langle y|r + h\rangle\right|^2 \notag \\
  & = \frac{1}{|G||H|}\left|\chi_y(r)\sum_{h\in H}
    \chi_y(h)\right|^2 \notag \\
    & = \frac{1}{|G||H|} \left|\chi_y(r)\chi_y(H)\right|^2,\label{eq:charsum}
\end{align}
where $\chi_y(H)$ denotes the unnormalized sum of elements.
Since $\chi_y(r)$ is a phase, it drops out, and the Fourier
sampling distribution is the same for any coset.

It remains to compute the sum of phases $\chi_y(H)$ 
in (\ref{eq:charsum}).
Using multiplicativity once more:
\begin{align}
  \chi_y(H)^2 & = \chi_y(H + H) \notag\\ & = |H| \chi_y(H) \notag \\ \Longrightarrow \quad
  0 & = \chi_y(H)(|H| - \chi_y(H)).\label{eq:charsum2}
\end{align}
Hence, $\chi_y(H) = 0$ or $\chi_y(H) = |H|$.
The former corresponds to perfect \emph{destructive} interference, and the
latter to \emph{constructive} interference.

\begin{example}
For $H = 2\mathbb{Z}_{12}$, we can evaluate character sums $\chi_y(H)$
visually, adding phases top-to-tail. For instance, for $\chi_1$, we
see that the phases cancel:
\vspace{-5pt}
\begin{center}
\includegraphics[scale=0.27]{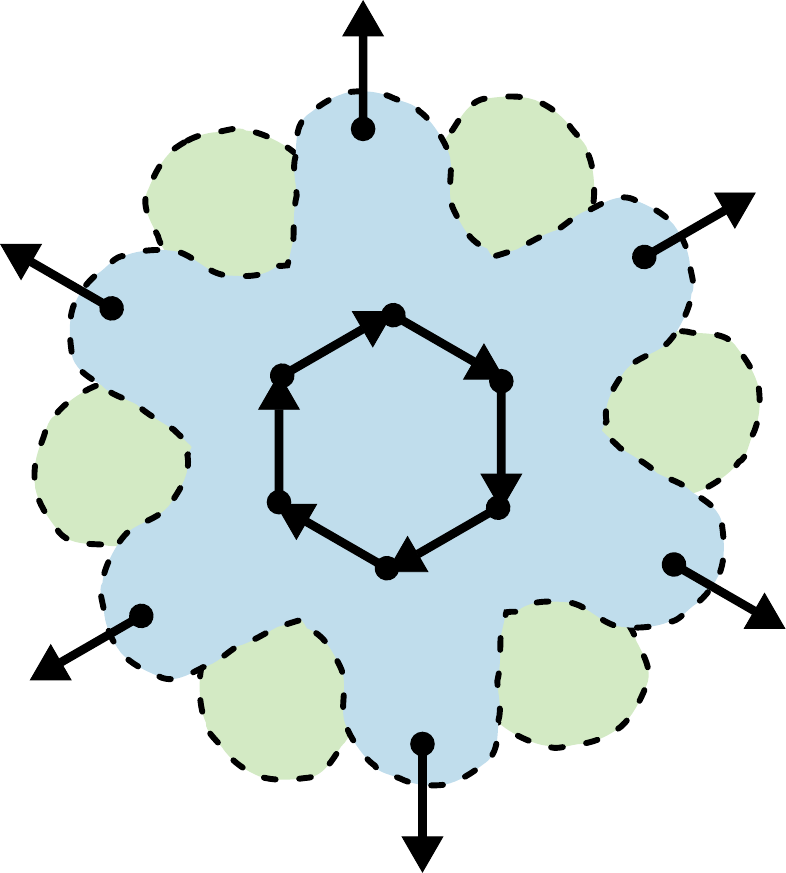}
\end{center}
\vspace{-5pt}
Thus, we have perfect destructive interference.
In general, $\chi_y(H) = 0$ for $2y$ not divisible by $12$.
\end{example}

As a result of interference, we only measure $\chi_y$ for which $\chi_y(H)=|H|$,
or equivalently, $\chi_y(h) = 1$ for all $h\in H$.
We say that $\chi_y$ \emph{annihilates} $H$, and call the set of
characters which annihilate $H$ the \emph{annihilator} $H^\perp
\subseteq \hat{G}$.
From (\ref{eq:charsum}), the probability of observing any $\hat{y}
\in H^\perp$ is $\hat{p}(y)=|H|/|G|$, and hence $|H^\perp| = |R| = |G|/|H|$.

If we observe $\hat{y}$, we know that the hidden subgroup is
annihilated by $\chi_y$. The \emph{kernel} $K_y$ of $\chi_y$ is the set of
elements assigned $1$:
\begin{equation}
  \label{eq:1}
  K_y = \{g\in G : \chi_y(g) = 1\}.
\end{equation}
Thus, measuring $\hat{y}$ implies that the hidden subgroup is contained in $K_y$. 
We prove in Appendix \ref{sec:kernel-intersection} that the intersection of $T=O(\log|G|)$ such kernels almost certainly equals the hidden subgroup.

\begin{example}
The annihilator of $H = 2\mathbb{Z}_{12}$ is $H^\perp = \{\chi_0,
\chi_6\}$, since these are the only characters constant on $H$:
\vspace{-10pt}
\begin{center}
\includegraphics[scale=0.26]{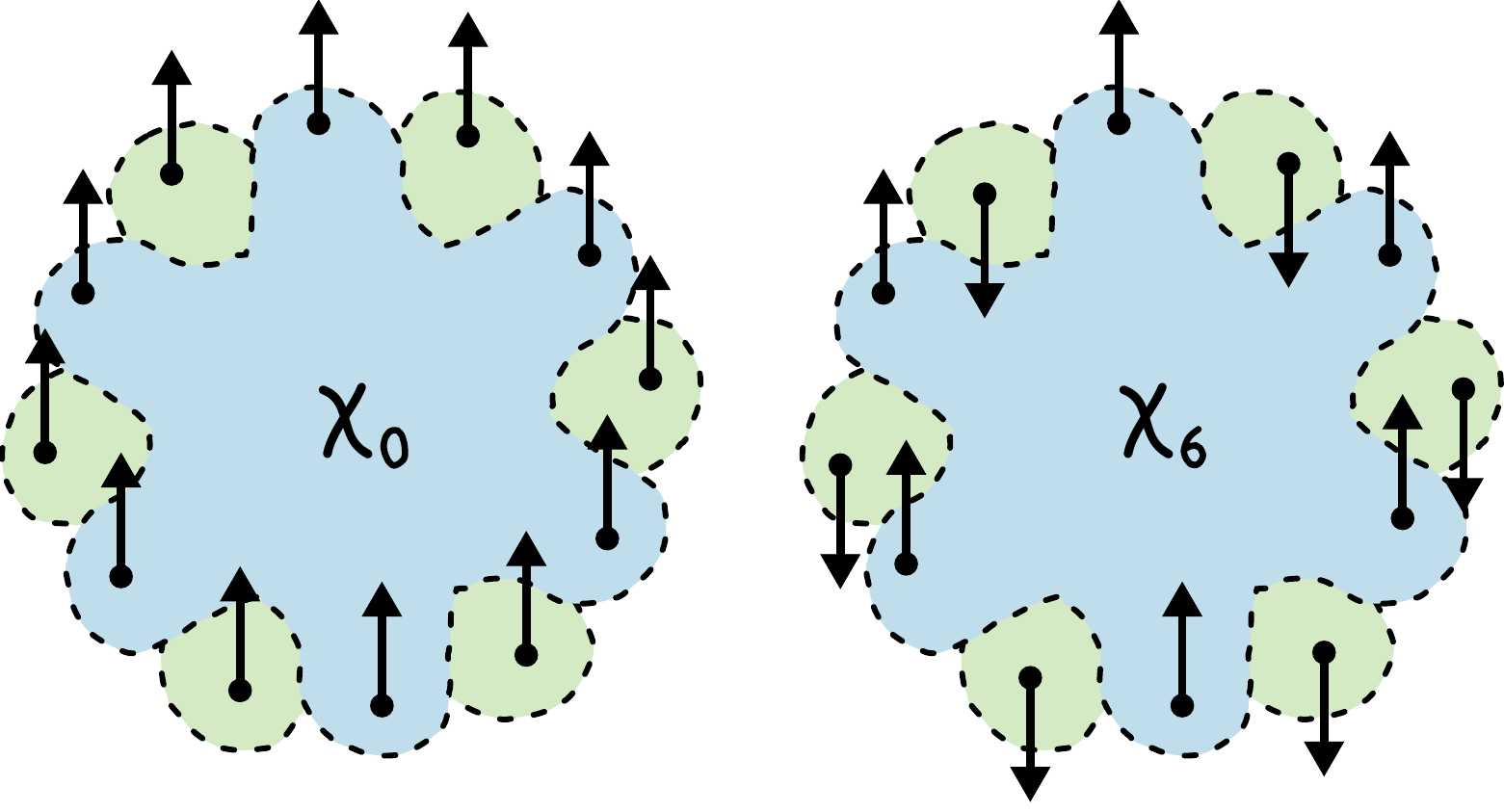}
\end{center}
\vspace{0pt}
The kernels are $K_0 = G$ and $K_2 = H$.
Measuring $\hat{0}$ and $\hat{6}$ is also consistent with $H' =
4\mathbb{Z}_{12}$, since
\[
(H')^\perp = \{\chi_0, \chi_3, \chi_6, \chi_9\}
\]
If we only observe $\hat{0}$ and $\hat{6}$, our running intersection
will be $H$; if we see either $\hat{3}$ or $\hat{9}$, we update the running intersection to $H'$.
\end{example}

\subsection{Information loss}
\label{sec:sampling-noise}

Let us now apply this algorithm in the learning scenario, where the oracle is replaced
by a training set $\mathcal{T} = \{(g, f(g))\}_{g\in X}$.
Quantum-mechanically, we assume black box access to a \emph{training state} analogous to (\ref{eq:r+H}):
\begin{align}
  \label{eq:psi_T}
  |\Psi_{\mathcal{T}}\rangle & = \frac{1}{\sqrt{N}} \sum_{g\in X} |g,
                   f(g)\rangle = \sum_{r\in R} \sqrt{\frac{|X_r|}{N}}|X_r, f(r)\rangle,
\end{align}
where $X_r = X \cap (r + H)$ is the set of training inputs lying in
coset $r + H$. We do not need to make any assumptions about the distribution from which the data is sampled, but work with the intuition of uniform sampling, where it is highly unlikely that two data samples are sampled from the same coset. 
Discarding the label gives a random \emph{partial coset
  state} $|X_r\rangle$ with probability $p_r = |X_r|/N$. Note that preparing such a state only takes time linear in the number of training data.

We might hope to infer $H$ by Fourier sampling as before. Unfortunately, for realistic data sets, which are small compared to the size of the group, this approach breaks down.
The problem is that the partial cosets $X_r$ are combined incoherently and no longer interfere with each other. This causes probability to ``leak" out of the annihilator of the hidden subgroup, as depicted in Fig.~\ref{fig:leak}. We prove in Appendix \ref{sec:leak} that this leakage is approximately linear in the size of the training set.
Thus, for realistic data sets, the signal gets drowned in noise.

As an extreme example, when the set of inputs $X = R$, corresponding to one data point per coset $x_r \in r + H$, the resulting probability distribution is uniform, since for each $r$, any character is equally likely to be sampled:
\[
  \hat{p}(y|r)= \left|\langle \hat{y}|x_r\rangle\right|^2 =
  \frac{1}{|G|}|\chi_y(x_r)|^2 = \frac{1}{|G|}.
\]
We get the same distribution for any set of coset representatives $R$, for any subgroup $H$, so the standard quantum algorithm is completely uninformative.

\begin{figure}[t]
  \centering
  \includegraphics[scale=0.3]{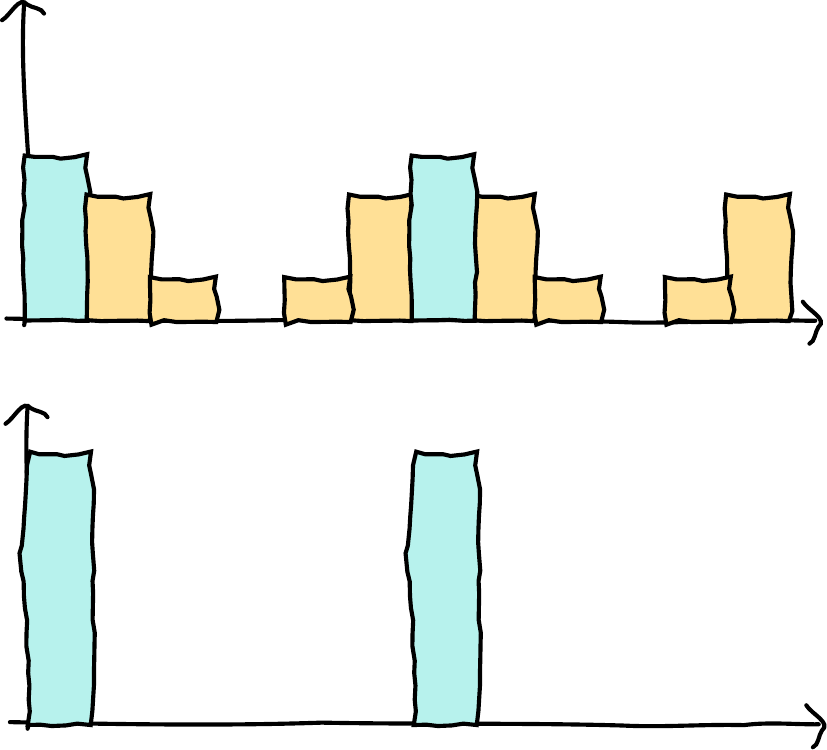}
  \caption{\emph{Top}. Since partial cosets do not interfere, probability ``leaks'' out of $H^\perp$. \emph{Bottom.} Measurement probability is concentrated on
    $H^\perp$ in the exact case.}
  \label{fig:leak}
\end{figure}

\section{Sample complexity}
\label{sec:sample}

Although the standard HSP quantum routine does not work on a small data  set, it may furnish the tools to construct an algorithm that does.
Before we attempt this construction, however, it is worth asking if a small training set contains enough information to reconstruct the hidden subgroup in principle. This is a necessary condition for the existence of learning algorithms.
Considering the strong symmetry of the problem, the answer is unsurprisingly affirmative: only a logarithmic number of data samples is needed to guess the hidden subgroup. As mentioned, we do not know if time-efficient algorithms exist to make use of this information, but will only look at heuristic principles below.

To compute the number of samples needed to learn the hidden subgroup, we need to make precise what we mean by ``learn''.
We will adopt the elegant framework of \emph{Probably Approximately
Correct (PAC) learning}, introduced for classical problems by
Valiant \cite{PAC} and extended to the quantum case by Bshouty and Jackson \cite{qPAC}.

\begin{figure}[h]
  \centering
  \includegraphics[scale=0.38]{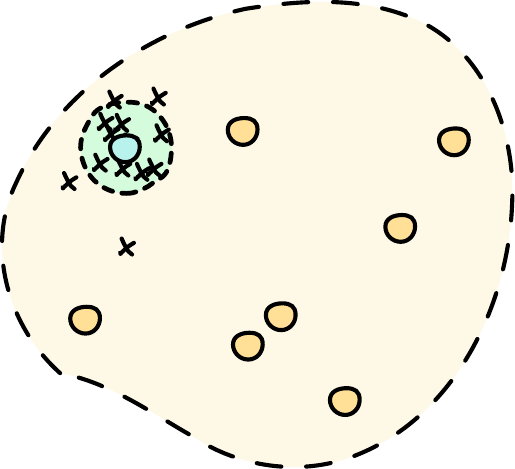}
  \caption{PAC: the blue dot is ground truth, yellow dots other concepts, and crosses runs of the algorithm.}
  \label{fig:pac}
\end{figure}

The PAC framework captures both the reliability of a learning
algorithm and the accuracy of the resulting estimate.
Informally, we say an algorithm is a PAC learner if it has a high
probability (``Probably'') of landing within a small neighbourhood of the correct
answer (``Approximately Correct''), as illustrated in Fig.~\ref{fig:pac}.

\vspace{3pt}
\vspace{-2pt}

Note that we restrict possible ground truths to lie in a
\emph{concept class}.
This is analogous to a prior
in Bayesian inference or a promise in the complexity literature. 
Indeed, in the learning version of the HSP, the subgroup promise
translates into the concept class. 

\subsection{Quantum PAC learning}
\label{sec:pac-learning}

With these intuitions in place, we can introduce the formal
definitions of PAC learning, both classical and quantum, and state
basic results about sample complexity in the PAC framework. This
involves a combinatorial parameter called the \textit{VC dimension}, which we
compute in the group-theoretic context in the subsequent section.

\subsubsection{Basic definitions}
\label{sec:basic-definitions}

Let $\mathcal{A}$ be a learning algorithm which uses examples to approximate the target function $f: \mathcal{X} \to \mathcal{Y}$.
We assume this function is deterministic, with $\mathcal{X}$ and $\mathcal{Y}$ finite for simplicity.
The target function is promised to be of the form $f_c: \mathcal{X} \to \mathcal{Y}$ for some $c \in
\mathsf{C}$, where $\mathsf{C}$ is the \emph{concept class}.
The outputs of the algorithm $\mathcal{A}$ are of the form $f_h:
\mathcal{X}\to \mathcal{Y}$, for
$h \in \mathsf{H}$, where $\mathsf{H}$ is the \emph{hypothesis class},
which may be distinct from the concept class.

Loosely speaking, an algorithm $\mathcal{A}$ PAC learns if, given enough
data, it reliably outputs an answer close to the target concept.
We quantify closeness with $\epsilon$ and reliability with $\delta$,
and say more precisely that $\mathcal{A}$ is a
$(\epsilon, \delta)$-\emph{PAC learner} if, for any target concept $c \in \mathsf{C}$
and distribution $\mathcal{D}$ over training data, with probability at least $1 -
\delta$, the algorithm outputs a guess $h\in \mathsf{H}$ which is
$\epsilon$-close to the ground truth, in the sense that
\begin{equation}
  \label{eq:pac-def}
  \mathbb{P}\left[\mathbb{P}_{x \sim \mathcal{D}}[f_{c}(x) \neq
  f_{h}(x)] \leq \epsilon\right] \geq 1 - \delta.
\end{equation}
If we think of our functions as labeling inputs, then ``approximately
correct'' means that our hypothesis disagrees with the true labeling with probability $\epsilon$, when inputs are drawn with the
same distribution as the training data.

For learning the HSP, we define the concept class as the set of all
possible subgroups of $G$:
\begin{equation}
  \label{eq:concept}
  \mathsf{C}_G = \{H \leq G\}.
\end{equation}
For the moment, we also take $\mathsf{H}_G = \mathsf{C}_G$.
As discussed above, our functions will not be the labels $f : G \to
S$, but rather the binary function $R_H: G \times G \to \{\texttt{0}, \texttt{1}\}$
which tells us when group elements are in the same coset.

\begin{example}
  For $G = \mathbb{Z}_{12}$, the concept class of cosets
  $\mathsf{C}_G$ corresponds to the factors of $12$:
\begin{center}
\includegraphics[width=0.32\columnwidth]{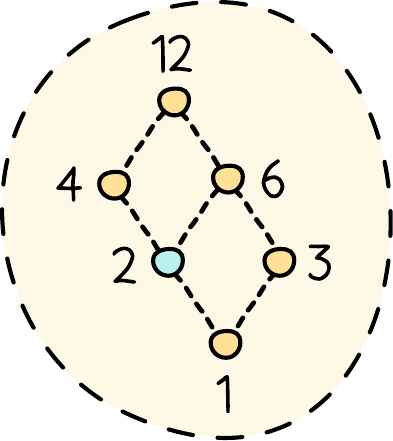}
\end{center}
An entry $c$ indicates a subgroup $H' = c\mathbb{Z}_{12}$.
\end{example}

\subsubsection{Samples and VC dimension}
\label{sec:vc-dimension}

We now turn to training data and sample complexity, and give a short summary about a fundamental result in quantum learning theory, which we adapt for the learning variant of the HSP in the next section.
In the classical case, we assume there is a black box which produces
random training pairs $(x, f_c(x)) \in \mathcal{X} \times \mathcal{Y}$ for a fixed, unknown
distribution $\mathcal{D}$ over $\mathcal{X}$. More formally, we have
a random \emph{example oracle} $\mathsf{EX}(c, \mathcal{D})$ which obeys
\begin{equation}
  \label{eq:16}
  \mathbb{P}[\mathsf{EX}(c, \mathcal{D}) = (x, f_c(x))] = \mathcal{D}(x).
\end{equation}
The sample complexity of a classical PAC learner is the number of
times $\mathsf{EX}$ must be invoked.

For quantum PAC learning, we can use the \emph{quantum example oracle}
introduced by Bshouty and Jackson \cite{qPAC}.
This encodes the probability distribution into the amplitudes of a
fixed quantum state:
\begin{equation}
  \label{eq:qex}
  \big|\mathsf{QEX}(c, {\mathcal{D}})\big\rangle =
  \sum_{x\in
    \mathcal{X}}\sqrt{\mathcal{D}(x)}|x, f_c(x)\rangle.
\end{equation}
Note that measurement in the $\mathcal{X}\times\mathcal{Y}$ basis yields a classical
example $\mathsf{EX}(c, \mathcal{D})$.
For more on the connection between the training state (\ref{eq:psi_T}) and example oracle (\ref{eq:qex}), see Appendix \ref{sec:sample-eq}.

\begin{figure}[t]
  \centering
  \includegraphics[scale=0.4]{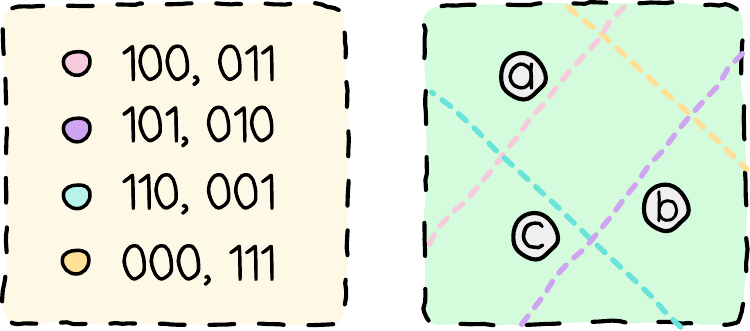}
  \caption{The concept class $\mathsf{C}$ (left)
    consists of four straight lines, while $\Gamma
    = \{a, b, c\}$ is represented by three points (right). Since lines can separate points in any desired way, $\mathsf{C}$ shatters $\Gamma$.}
  \label{fig:shatter}
\end{figure}

A fundamental result in statistical learning theory is that the
number of samples needed for PAC learning is given by the \emph{Vapnik-Chervonenkis (VC)
dimension} \cite{Vapnik2015} of the concept class.
To define VC dimension, we first need the concept of \emph{shattering}.
A set of inputs $\Gamma \subseteq \mathcal{X}$ is
\emph{shattered} by $\mathsf{C}$ if, for every possible assignment of
labels in $\mathcal{Y}$ to elements of $\Gamma$, there is some $f_c$
that realizes this assignment.
Letting $\mathcal{Y}^\Gamma$ denote the set of maps
$\Gamma\to\mathcal{Y}$, we can write the shattering condition as
\begin{equation}
  \label{eq:18}
\mathcal{Y}^\Gamma =  \{f_c |_\Gamma : c \in \mathsf{C}\} = f_{\mathsf{C}}|_\Gamma,
\end{equation}
where the last expression is shorthand for the middle.
We picture this for binary classification ($\mathcal{Y} =
\{0, 1\}$) of three points on a plane in Fig.~\ref{fig:shatter}.

The VC dimension of $\mathsf{C}$ is the cardinality of the largest set
of points $\Gamma$ that it can shatter:
\begin{equation}
  \label{eq:19}
  \text{dim}_\text{VC}(\mathsf{C}) = \max \{|\Gamma| : \Gamma
  \subseteq \mathcal{X}, \, \mathcal{Y}^\Gamma =  f_{\mathsf{C}}|_\Gamma\}.
\end{equation}
Next to each concept in Fig.~\ref{fig:shatter} (left), we've given the
labels it assigns, in binary, for $a$, $b$ and $c$; the first string
is $0$ is assigned below the line, the second above.
We see that the VC dimension corresponds to a \emph{binary description
  length} of the concept in terms of examples.

It is no surprise, then, that the VC dimension is related to the number of examples we need to specify the concept.
What is much more surprising is that this number is exactly the same
for both quantum and classical learning.
It can be shown for the classical \cite{Blumer, hanneke2016optimal}
and quantum case \cite{Atici_2005,
  ZHANG201040, arunachalam2017optimal} that the sample complexity of
an $(\epsilon, \delta)$-PAC learner with VC dimension
$\text{dim}_\text{VC}(\mathsf{C}) = d$ is 
\begin{equation}
  \label{eq:pac-sample}
  N = \Theta \left( \frac{d}{\epsilon} + \frac{\log \delta^{-1}}{\epsilon}\right),
\end{equation}
Thus, from the sample complexity perspective, classical and
quantum PAC learning are equivalent.

In general, computing the VC dimension is a combinatorially formidable
task, and for many concept classes it is unknown.
Luckily, with a structural conjecture about abelian groups, we will
be able to calculate the VC dimension for our problem explicitly.

\begin{example}
For $G = \mathbb{Z}_{12}$, consider binary training examples
  \[
    \Gamma = \{(0, 3), (3, 5)\}.
  \]
  This is shattered by subgroups of $G$. We denote a label $\texttt{0}$ by a solid line (indicating two integers are in one coset) and labbel $\texttt{1}$ by a dashed line (indicating they are not):
  \vspace{2pt}
\begin{center}
\includegraphics[scale=0.27]{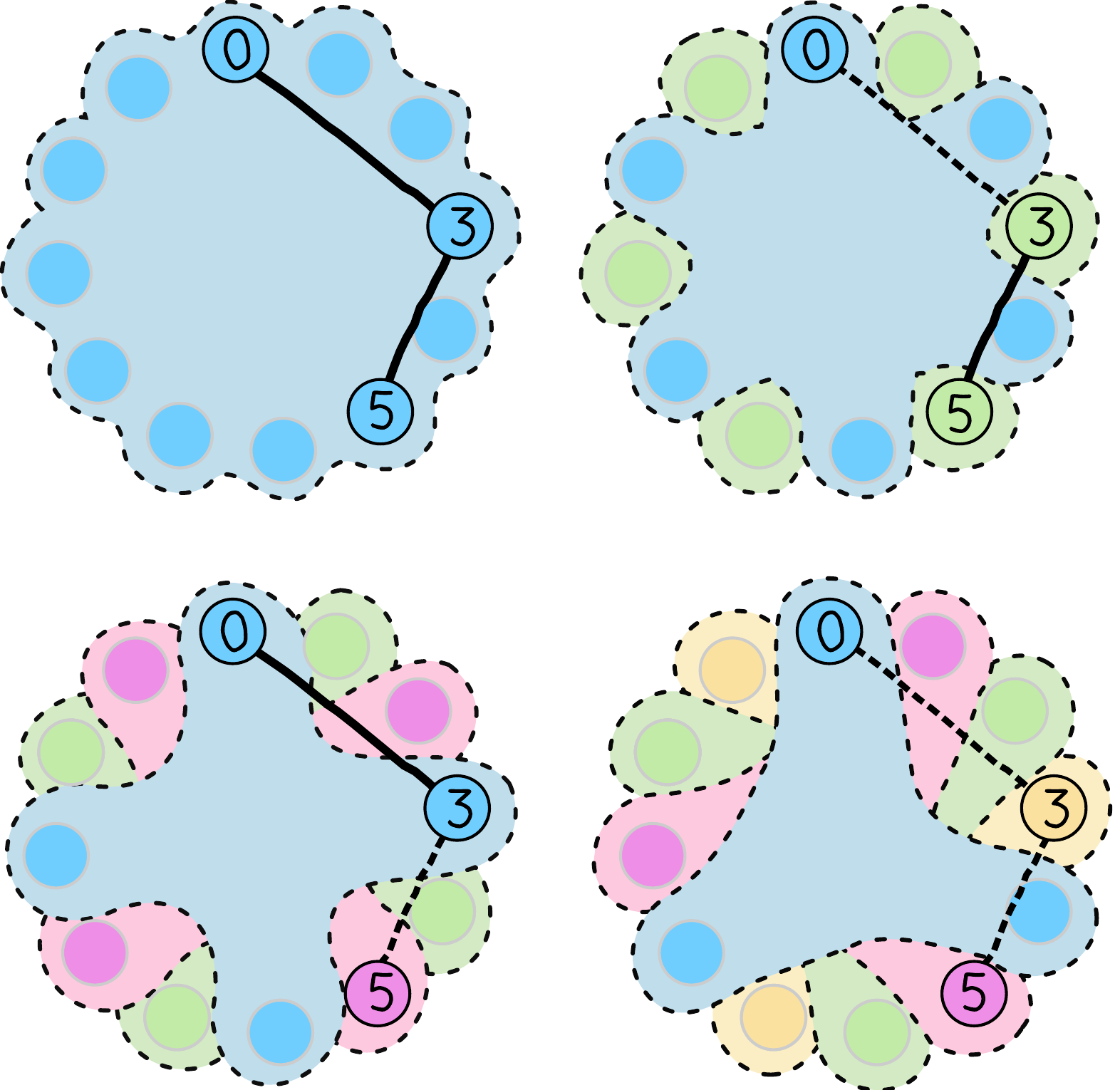}
\end{center}
\vspace{-0pt}
We can then label subgroups with the two bits they assign to the elements of $\Gamma$:
  \vspace{2pt}
  \begin{center}
\includegraphics[width=0.32\columnwidth]{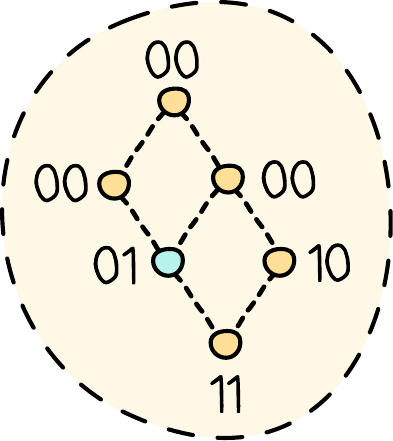}
\end{center}
  \vspace{2pt}
Since there are only $6 < 2^3$ subgroups, $\mathsf{C}_G$ cannot shatter a
set of size $\geq 3$. Thus, $\text{dim}_\text{VC}(\mathsf{C}_G) = 2$.
\label{ex:VC}
\end{example}

\subsection{QPAC for abelian groups}
\label{sec:struct-abel-groups}

Let us now return to the HSP.
Recall that our concept class $\mathsf{C}_G =
\{H \leq G\}$ is the set of subgroups of $G$, and the functions of
interest are the binary relations $R_H$ induced by these subgroups.
To find the sample complexity,
we need to compute the
VC dimension of $\mathsf{C}_G$.

To make progress, we need more group theory.
Using the terminology of Appendix \ref{sec:group-theory}, we can state the \emph{fundamental theorem of finite abelian
  groups}: every finite abelian group $G$ is
isomorphic to a direct sum of cyclic groups,
\begin{equation}
  \label{eq:ftfag}
  G \cong \mathbb{Z}_{q_1}^{\ell_1} \oplus \mathbb{Z}_{q_2}^{\ell_2} \oplus \cdots \oplus \mathbb{Z}_{q_M}^{\ell_M},
\end{equation}
where the $q_i = p_i^{m_i}$ are distinct prime powers, $\ell_i \in \mathbb{N}$, and $|G| =
\prod_{i=1}^M \ell_iq_i$.
Moreover, this decomposition is unique up to the order of factors.
A proof of (\ref{eq:ftfag}) can be found in any good algebra textbook, 
e.g. \cite{algebra, lang02, Dummit1999AbstractA}.

In Appendix \ref{sec:sample-eq}, we show that a cyclic group of prime
power order has VC dimension $1$, and that (subject to technical conjecture) VC dimension is
additive for the direct sum of such factors.
It follows that, for the decomposition (\ref{eq:ftfag}), the VC
dimension depends on the total number of cyclic factors.
Thus, we have:

\begin{theorem}
  \label{thm:pac}
  For an abelian group $G$ with decomposition (\ref{eq:ftfag}),
  $(\epsilon, \delta)$-PAC learning the hidden subgroup requires
  \begin{equation}
    N_{\mathrm{bin}} = \Theta \left( \frac{1}{\epsilon}\sum_{i=1}^M\ell_i + \frac{\log
        \delta^{-1}}{\epsilon}\right)\label{eq:Nbin}
  \end{equation}
binary examples for classical or quantum algorithms.
\end{theorem}
\noindent A training set $\mathcal{T}$ of $N$ labeled examples
is equivalent to a set of $N_\text{bin} \approx N^2/2$ binary examples, so we expect
a sample complexity of $N = \Theta(\sqrt{N_\text{bin}})$ labeled examples
to learn the hidden subgroup.

\vspace{4pt}
\begin{figure}[h]
  \centering
  \includegraphics[width=0.64\columnwidth]{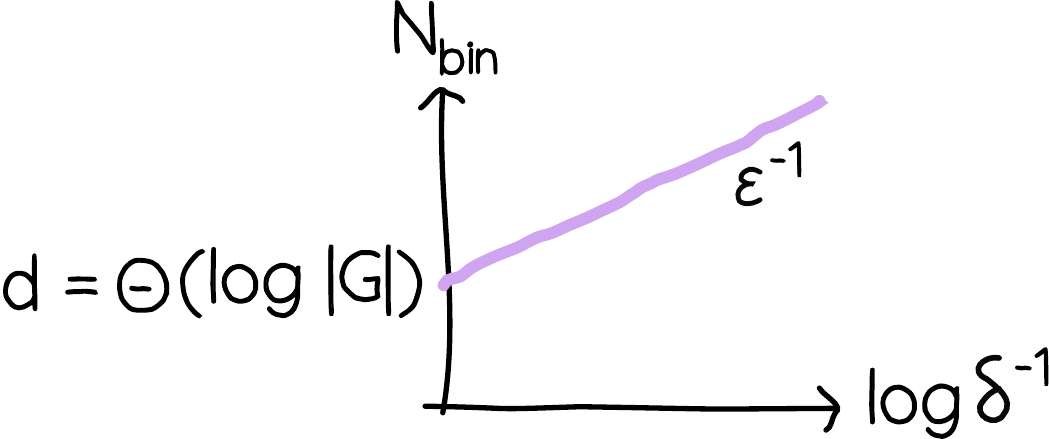}
  \caption{The sample complexity is a linear function of $\log \delta^{-1}$, with slope $\epsilon^{-1}$ and intercept $\Theta(\log|G|)$.}
  \label{fig:pac-thm}
\end{figure}
\vspace{-3pt}

This is a logarithmic number of observations in the sense that when $G
= \mathbb{Z}_{q}^\ell$ for fixed $q$, or a product of factors
(\ref{eq:ftfag}) with $q_i = O(1)$, the sample complexity is
$\sum_i\ell_i = \Theta(\log |G|)$. 
We picture this scaling in Fig. \ref{fig:pac-thm}.
Note that, when we learn with \emph{specific} training
distributions, (\ref{eq:Nbin}) will only 
upper bound sample complexity.

\section{Inference with quantum computers}
\label{sec:reveal}

As we saw in the previous section, the data samples given as input to the learning variant of the Hidden Subgroup Problem may contain the information we need in order to identify the hidden subgroup. That is what sample complexity captures. But the fact they contain this information does not provide a way to extract it.

In this section, however, we show how the standard HSP routine provides intuition to construct  heuristic quantum algorithms for learning the hidden subgroup from data.
We start by presenting and motivating a general inference principle based on the standard algorithm, and then convert this into a cost function we can formally analyze and implement as a quantum circuit.

\subsection{Inference from invariance}
The standard HSP algorithm converts coset states $|r+H\rangle$ from the oracle, to states spanned by characters $|\chi\rangle$ from the annihilator, $\chi\in H^\perp$. As we will explain below, these states form an invariant subspace under transformations from the hidden subgroup.
When the oracle generating the coset states is replaced by training data, information leaks, and the ``data states'' $|X_r\rangle$ can no longer be expanded in the annihilator $\chi\in H^\perp$ which best explains the data. This suggests the following principle of inference:
\begin{quote}
    \textit{Find the annihilator (subspace) which is closest to the data (subspace).}
\end{quote}
We call this ``inference to the nearest annihilator,'' or ``inference to the nearest invariant subspace". 

To flesh this out, let us define the subspaces involved.
The \emph{data subspace} $\mathcal{H}_{\mathcal{T}}$ is the space spanned by partial coset states $|X_r\rangle$, with states of the form
\[
|\psi\rangle = \sum_{r\in R} \alpha_r |X_r\rangle,
\]
where the $\alpha_r \in \mathbb{C}$ are amplitudes.
The \emph{annihilator subspace} $\mathcal{H}_{\tilde{H}^\perp}$ is spanned by character states $|\chi\rangle$ 
from the annihilator $\tilde{H}^\perp$ and contains states
\[
|\phi\rangle = \sum_{\tilde{\chi}\in\tilde{H}^\perp} \alpha_{\tilde{\chi}} |\tilde{\chi}\rangle = \sum_{g\in G} c_g |g\rangle,
\] 
with unconstrained amplitudes $\alpha_{\chi}$ and computational basis coefficients $c_g = \langle g|\phi\rangle$.
As discussed in Appendix \ref{sec:QFT}, the character states $|\tilde{\chi}\rangle$ for $\tilde{\chi}\in \tilde{H}^\perp$ span precisely the joint $1$-eigenspace of the operators mapping $|g\rangle \mapsto |g+\tilde{h}\rangle$ for $\tilde{h}\in\tilde{H}$, or equivalently, shifting coefficients $c_g \mapsto c_{g-\tilde{h}}$.
This means that the annihilator subspace $\mathcal{H}_{\tilde{H}^\perp}$ consists of all states
such that, for any $\tilde{h}\in\tilde{H}$, $c_g = c_{g-\tilde{h}}$.

\begin{example}
Consider a data set $\mathcal{T} = \{(0,
  \textcolor{cyan}{\bullet}), (2, \textcolor{cyan}{\bullet}), (3,
  \textcolor{lime}{\bullet})\}$:
\begin{center}
\includegraphics[scale=0.27]{pics/oracle9}
\end{center}
  The partial cosets are $X_{\textcolor{cyan}{\bullet}} = \{0, 2\}$ and $X_{\textcolor{lime}{\bullet}} = \{3\}$, so $\mathcal{H}_\mathcal{T}$ consists of states $\alpha_{\textcolor{cyan}{\bullet}}|X_{\textcolor{cyan}{\bullet}}\rangle + \alpha_{\textcolor{lime}{\bullet}}|X_{\textcolor{lime}{\bullet}}\rangle$
  for all $|\alpha_{\textcolor{cyan}{\bullet}}|^2+|\alpha_{\textcolor{lime}{\bullet}}|^2=1$.
  The subgroup $\tilde{H}=3\mathbb{Z}_{12}$ has annihilator $\tilde{H}^\perp = \{\chi_0, \chi_4, \chi_8\}$, with subspace $\mathcal{H}_{\tilde{H}^\perp}$ consisting of states
  $\hat{\alpha}_0|\hat{0}\rangle+\hat{\alpha}_4|\hat{4}\rangle+\hat{\alpha}_8|\hat{8}\rangle$ for all amplitudes $|\hat{\alpha}_{0}|^2+|\hat{\alpha}_{4}|^2+|\hat{\alpha}_{8}|^2=1$.
\end{example}

Our proposed inference principle compares the data to the annihiliator subspace; for the moment we do not specify according to which metric.
To understand the motivation for this idea, we can recast the standard HSP routine as a simple instance of this principle.
Instead of the data subspace, we have the \emph{oracle subspace} $\mathcal{H}_{f}$ that represents the full data distribution. This space is spanned by the coset states $|r + H\rangle$, with states
\[
|\psi\rangle = \sum_r \alpha_r |r + H\rangle
\]
for amplitudes $\alpha_r$.

\vspace{0pt}
\begin{figure}[t]
  \centering
  \includegraphics[scale=0.33]{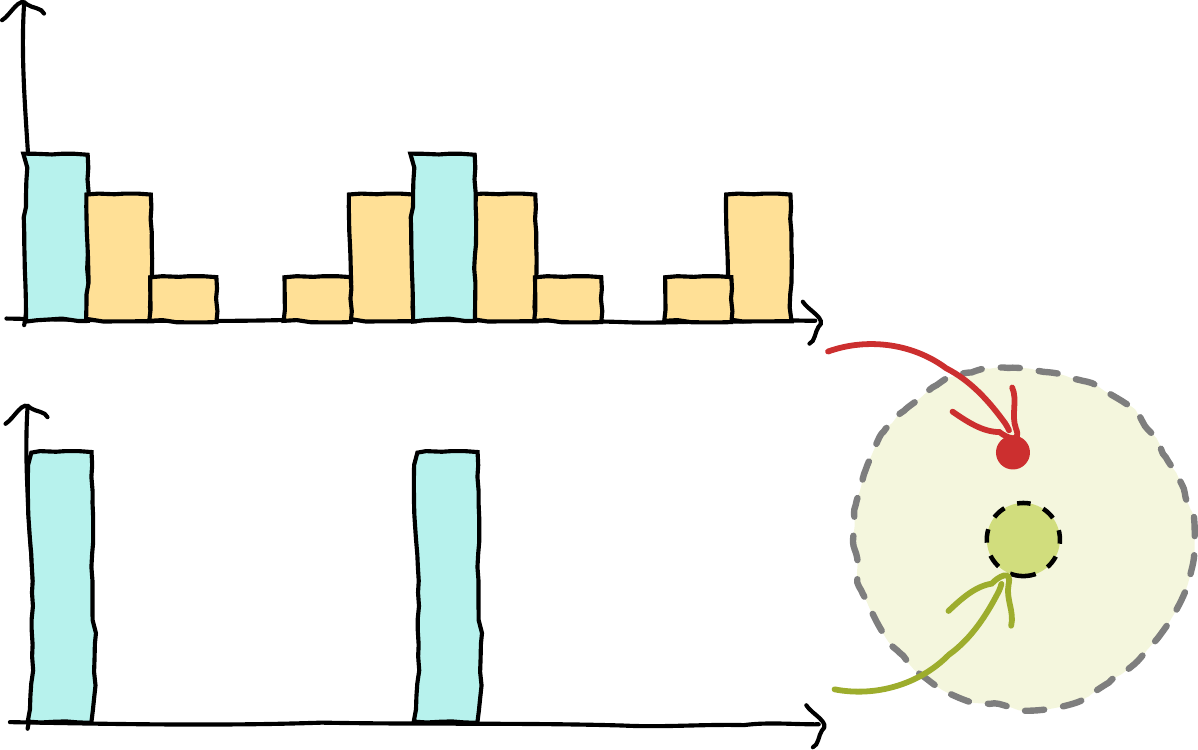}
  \caption{The data subspace (red) can be viewed as an annihilator plus error, and ``corrected" to the nearest annihilator subspace (green).}
  \label{fig:error}
\end{figure}
\vspace{0pt}

Since $H + h = H$ for any element $h \in H$, these states are invariant under shifts $|g\rangle \mapsto |g+h\rangle$ for any element of $H$. But, as we argued above, any states fixed by all such shifts must lie in the annihilator subspace, so $\mathcal{H}_f \subseteq \mathcal{H}_{H}^\perp$.
Finally, note that both subspaces have the same dimension. There are $|G|/|H| = |H^\perp|$ cosets as well as characters in the annihilator. It follows that the two are equal, $\mathcal{H}_{f} = \mathcal{H}_{H^\perp}$, and the standard HSP algorithm produces an annihilator subspace. We sample from this subspace in the Fourier basis to reveal its identity.

With access to only a few data samples, guessing the hidden subgroup becomes an inference or ``learning'' task.  The data subspace $\mathcal{H}_\mathcal{T}$ will no longer exactly equal the annihilator subspace $\mathcal{H}_{\tilde{H}^\perp}$ of any subgroup $\tilde{H}$ due to information leakage, as discussed above.
But as shown in Appendix \ref{sec:leak}, for sufficient data, the true annihilator $H^\perp$ remains a better explanation of the weak signal than any other candidate $\tilde{H}^\perp$. We picture this in Fig.~\ref{fig:error}.

\subsection{DAOism}
\label{sec:fourier-cost}

Our goal now is to find ways to compare the data subspace $\mathcal{H}_{\mathcal{T}}$ to candidate annihilator subspaces $\mathcal{H}_{\tilde{H}^\perp}$.
There are many ways to compute a distance between subspaces \cite{ye2016}, but these tend to be computationally prohibitive. We want to find a distance measure, or perhaps a surrogate, which we can easily compute.

A simple strategy is to define a data state and a canonical state in $\mathcal{H}_{\tilde{H}^\perp}$ and maximize their overlap.
After appropriate regularization, this gives a well-behaved cost function we call the \emph{data-annihilator overlap (DAO)}. Optimizing this gives the ``closest" annihilator, and hence our best guess at the subgroup that generated the data according to our proposed principle of inference.
Below, we give a precise definition, analyze its inductive bias, and show how to implement it on a quantum computer.

\subsubsection{Defining the DAO}

To compute the DAO, we need to encode our data into a mixture of partial coset states
\begin{align}
    \rho_\mathcal{T} = \frac{1}{N}\sum_{r\in R} |X_r| |X_r\rangle\langle X_r|. \label{eq:rho_T}
\end{align}
Concretely, we could obtain this from the pure state $|\Psi_\mathcal{T}\rangle$ from (\ref{eq:r+H}), where group elements $g$ are entangled with their labels $f(g)$, and the label register is traced out.

A natural choice of state in the annihilator subspace is the uniform superposition of characters:
\begin{equation}
    |\tilde{H}^\perp\rangle = \frac{1}{\sqrt{|\tilde{H}^\perp|}}\sum_{\hat{y}\in \tilde{H}^\perp} |\hat{y}\rangle.
\end{equation}
This is the result of applying the inverse QFT to the subgroup state $|H\rangle$,
since as the HSP algorithm shows,
$\langle y|\mathcal{F}^\dagger | \tilde{H}\rangle \propto 
\mathbb{I}[\hat{y} \in \tilde{H}^\perp]$.
It follows that the inverse QFT takes $H$ to its annihilator, $\mathcal{F}^\dagger |\tilde{H}\rangle= |\tilde{H}^\perp\rangle$.
We can therefore produce annihilator states either by direct oracle access to $|\tilde{H}^\perp\rangle$ or access to $|\tilde{H}\rangle$.

\begin{figure}[h]
  \centering
  \includegraphics[scale=0.32]{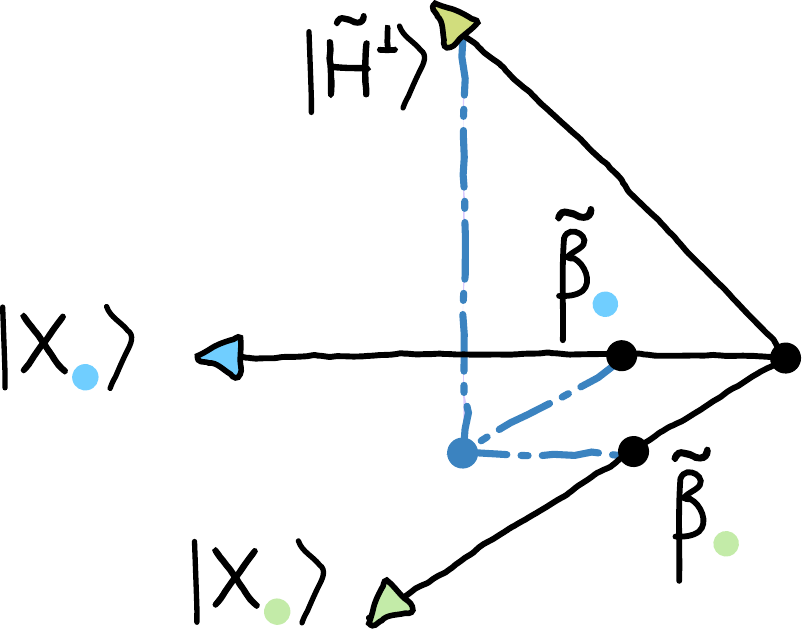}
  \caption{The DAO as coordinates for the annihilator
    $|\tilde{H}^\perp\rangle$ in the subspace spanned by $|X_r\rangle$.}
  \label{fig:dao}
\end{figure}

Since $\rho_\mathcal{T}$ is mixed, we can maximize squared fidelity rather than overlap between the training state and the annihilator state:
\begin{align}
\langle \tilde{H}^\perp|\rho_\mathcal{T}|\tilde{H}^\perp\rangle
& = \sum_{r\in R}\frac{|X_r|}{N} |\langle X_r|\tilde{H}^\perp\rangle|^2  = \Vert \beta(\tilde{H})\Vert_2^2,\label{eq:dao}
\end{align}
where we define the \emph{data-annihilator overlap (DAO) vector} $\beta(\tilde{H})$ by its components $\beta_r(\tilde{H})=\sqrt{|X_r|N^{-1}}\langle X_r|\tilde{H}^\perp\rangle$.
We think of $\beta(\tilde{H})$ as the projection of $|\tilde{H}^\perp\rangle$ onto the data subspace, with basis vectors $|X_r\rangle$ weighted by $\sqrt{|X_r|}$; see Fig.~\ref{fig:dao}.

\vspace{5pt}
\vspace{-0pt}

\begin{example}
Consider $\mathcal{T} = \{(0,
  \textcolor{cyan}{\bullet}), (2, \textcolor{cyan}{\bullet}), (3,
  \textcolor{lime}{\bullet})\}$, 
  $X_{\textcolor{cyan}{\bullet}} = \{0, 2\}$, $X_{\textcolor{lime}{\bullet}} = \{3\}$, and a candidate subgroup $\tilde{H}=3\mathbb{Z}_{12}$ with annihilator $\tilde{H}^\perp=\{\chi_0,\chi_4,\chi_8\}$. The DAO vector has components
  \begin{align*}
      \beta_{\textcolor{cyan}{\bullet}}(\tilde{H}) &= \sqrt{\frac{|X_{\textcolor{cyan}{\bullet}}|}{3}}\langle X_{\textcolor{cyan}{\bullet}}|\tilde{H}^\perp\rangle 
       = \frac{1}{2\sqrt{3}} \\
      \beta_{\textcolor{lime}{\bullet}}(\tilde{H}) &= \sqrt{\frac{|X_{\textcolor{lime}{\bullet}}|}{3}}\langle X_{\textcolor{lime}{\bullet}}|\tilde{H}^\perp\rangle
       = \frac{1}{\sqrt{3}}.
  \end{align*}
  The DAO vector has squared length
  \[
  \Vert \beta(\tilde{H})\Vert_2^2 = \left(\frac{1}{2\sqrt{3}} \right)^2 + \left(\frac{1}{\sqrt{3}} \right)^2 = \frac{5}{12}.
  \]
\end{example}

Conceptually, 
we can think of the DAO as how well the hypothesis $|\tilde{H}^\perp\rangle$ explains the data, with more weight accorded to larger clusters. Although this is a rather crude measure of distance between the data and annihilator and subspace, its main virtues are that (a) it has a simple interpretation, and (b) it is easily computed given access to candidate subgroups or annihilators. 

To practically use this as a cost function, we need first of all to determine an optimization schedule for evaluating subgroups. This is technically challenging problem beyond the modest scope of this paper.
A second problem is overfitting, since replacing $\tilde{H}^\perp$ with a larger annihilator $\tilde{J}^\perp \supseteq \tilde{H}^\perp$ automatically results in a larger overlap.
A simple cure is to penalize the size of the candidate annihilator $\tilde{H}^\perp$.
Based on these observations, we propose to use the \emph{DAO cost}:
\begin{align}
  \mathcal{C}_\text{~\!\customyinyang[1.2]}(\tilde{H})^2 & = -\Vert \beta(\tilde{H})
                                   \Vert_2 + \lambda |\tilde{H}^\perp| ,  \label{eq:dao-reg}
\end{align}
where $\lambda \geq 0$ is a regularization constant. 
The first term measures how well the data is explained by the
hypothesis, and the second, how economically it explains it.

\subsubsection{Consistency and bias}
\label{sec:dcota-bias}

The first question we should ask of any cost function is whether it gives the right answer for sufficiently large training sets
, a property termed ``consistency'' in machine learning.
For complete data, $X = G$, one can show that (\ref{eq:dao-reg}) becomes
\begin{align}
  \mathcal{C}_\text{~\!\customyinyang[1.3]}(\tilde{H})
                                  & = - \frac{|\tilde{H}_\cap^\perp|}{|R|}|\hat{R}\cap \tilde{H}_\cap|^{1/2} - \lambda |\tilde{H}^\perp|, \label{eq:full-1}
\end{align}
where $\tilde{H}_\cap^\perp = \tilde{H}^\perp\cap H^\perp$ is the intersection
of the true and candidate annihilators, $\tilde{H}_\cap= (\tilde{H}_\cap^\perp)^\perp$, and finally $\hat{R} = \{\hat{r} : r\in R\}$ are the characters labelled by the coset representatives $R$.

To analyse this expression, we can simplify it using a probabilistic approximation. 
When we label a coset $r + H$ with a representative, we can use any element; $r$ is an arbitrary choice.
We will therefore make the assumption that $\hat{R}$ is distributed \emph{randomly} in the ambient group,\footnote{Note that this assumption can be made rigorous by choosing uniformly at random within cosets and results from probabilistic number theory, but we defer this to future work.} in the sense that
\begin{equation}
    |\hat{R} \cap X| = \frac{|R||X|}{|G|} + O(|X|^{-1}). \label{eq:rand-sampler}
\end{equation}
Using $|\tilde{H}_\cap|=|G|/|\tilde{H}_\cap^\perp|$, (\ref{eq:full-1}) then simplifies to
\begin{align*}
    \mathcal{C}_\text{~\!\customyinyang[1.3]}(\tilde{H}) = -\sqrt{\frac{|\tilde{H}^\perp_\cap|}{|G|}} - \lambda |\tilde{H}^\perp| - O(|\tilde{H}_\cap^\perp|^{-1}). \label{eq:full-2}
\end{align*}
If $\lambda$ is large enough to dominate the $O(|\tilde{H}_\cap^\perp|^{-1})$ term, but small enough to be dominated by the DAO length, then the cost is minimized by first maximizing $|\tilde{H}^\perp_\cap|$, i.e. the overlap between the 
candidate annihilator $\tilde{H}^\perp$ and true annihilator $H^\perp$.

The annihilators of maximal overlap contain $H^\perp$:
\[
\tilde{H}^\perp\cap H^\perp = H^\perp \quad \Longrightarrow \quad
  H^\perp  \subseteq \tilde{H}^\perp.
\]
The smallest such annihilator is $H^\perp$ itself, so the regularization ensures that (subject to our randomness assumptions) the DAO cost is minimized by $H_\text{min} = H$, and is therefore consistent.

\begin{example}
Consider complete data for the hidden subgroup $H=2\mathbb{Z}_{12}$ of $G=\mathbb{Z}_{12}$, pictured as follows:
\begin{center}
\includegraphics[scale=0.27]{pics/oracle7}
\end{center}
Below, we list the different candidate subgroups $\tilde{H}$ of $G$, the size of the corresponding annihilator $|\tilde{H}^\perp|$, and the DAO length $\Vert \beta(\tilde{H})\Vert_2$:
\begin{center}
\begin{tabular}{ c | c  c c c c c}
$H$& $12\mathbb{Z}_{12}$& $6\mathbb{Z}_{12}$ &
                                                                  $4\mathbb{Z}_{12}$
  & $3\mathbb{Z}_{12}$ & $2\mathbb{Z}_{12}$ & $\mathbb{Z}_{12}$ \\  \hline
$|\tilde{H}^\perp|$&1&2&3&4&6&12
\\  
$\Vert \beta(\tilde{H})\Vert_2$&$\frac{1}{2\sqrt{3}}$&$\frac{1}{\sqrt{6}}$&$\frac{1}{2}$&$\frac{1}{\sqrt{6}}$&$\frac{1}{\sqrt{2}}$&$\frac{1}{\sqrt{2}}$
\end{tabular}
\end{center}
The true hidden subgroup $H =2\mathbb{Z}_{12}$ and full group have the smallest DAO length and explain the data equally well, but a small regularization term ensures $H$ has lowest DAO cost.
\end{example}

We can use similar techniques to understand the \textit{bias} of our proposed cost function when data is sparse: given multiple candidate subgroups that are compatible with the data, which one does the cost prefer?
We relegate details to Appendix \ref{sec:cost-funct-deta}, but the basic idea is to split a candidate annihilator $\tilde{H}^\perp$ into a part $\tilde{H}^\perp_\cap$ that overlaps with
$H^\perp$, and a residual part $\tilde{H}^\perp_0 =
\tilde{H}^\perp\backslash \tilde{H}^\perp_\cap$ that does not.
The characters in the first set will interfere constructively, and the
second destructively, resulting in different contributions to the cost function.

The constructive term is similar to the full data case, in the sense that cost is minimized by maximizing the overlap between $\tilde{H}^\perp$ and $H^\perp$. The remaining terms involve randomly fluctuating phase sums which average to zero with enough data. 
We argue in Appendix \ref{sec:cost-funct-deta} 
that these sums are suppressed by powers of $|R|/N$.\footnote{This is not quite the same as our sample complexity result, since it is hard to meaningfully establish the ``typical" size of $R$. We leave a detailed study of their relationship to future work.} For subgroups whose annihilators overlap $H^\perp$ equally, and which therefore have the same constructive term, the corrections tend to reward large $|\tilde{H}_0^\perp|$, and hence, larger annihilators $\tilde{H}^\perp$. We can use our regularization term to counteract this unwanted bias.

\subsubsection{Evaluation}

Our next task will be to provide a quantum circuit to
evaluate the DAO cost.
As discussed above, we assume we have an oracle encoding our training data in reduced form (\ref{eq:rho_T}), and
and a family of oracles which output a subgroup state, $\mathcal{O}_{\tilde{H}}|0\rangle
=|\tilde{H}\rangle$.\footnote{The cost of invoking such an oracle is the number of steps needed to specify $H$. We will not dwell on these subtleties here.}
Applying the inverse QFT to the state $|\tilde{H}\rangle$ yields the
annihilator state $|\tilde{H}^\perp\rangle$.
From (\ref{eq:dao}), we need an algorithm to compute fidelity.

\vspace{0pt}
\begin{figure}[t]
  \centering
  \includegraphics[scale=0.32]{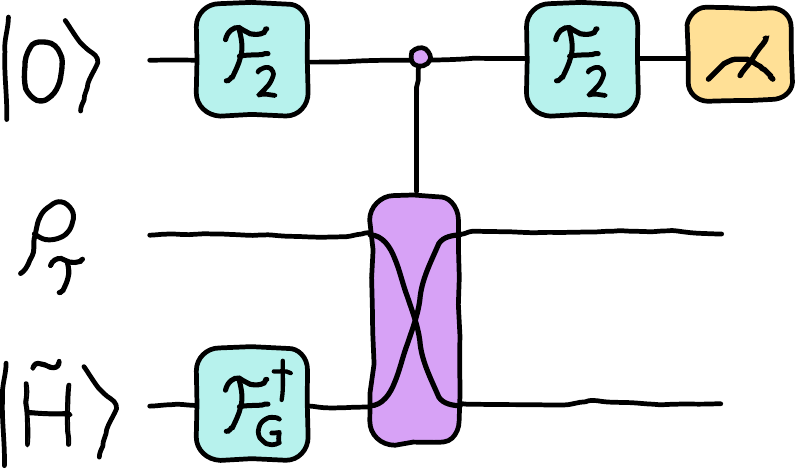}
  \caption{A SWAP circuit for evaluating the DAO.}
  \label{fig:swap}
\end{figure}
\vspace{0pt}

A simple approach is to use the SWAP test \cite{Buhrman_2001}.
The circuit Fig.~\ref{fig:swap} performs a
controlled swap of our two
states with an ancilla qubit (conjugated by Hadamards)
which it then measures.
This yields a measured bit $b \in \{0, 1\}$ with probability
\begin{equation}
  \label{eq:p_b}
    p_{b} = \frac{1}{2}\left[1 + (-1)^b \langle
                                                \tilde{H}^\perp |
                                                \rho_\mathcal{T}|\tilde{H}^\perp\rangle
                                              \right]. 
\end{equation}
We can use this to estimate the squared fidelity to additive error $\eta$ with
$O(1/\eta^2)$ trials.

Since estimating (\ref{eq:dao}) involves finding the correlation between a function (the indicator function on $X_r$) and the Fourier transform of a function (the indicator on $\tilde{H}$), computing the DAO can presumably be cast in the form of a forrelation problem \cite{aaronson2009bqp, aaronson2014forrelation}. One might therefore hope the quantum evaluation procedure exhibits a provable advantage in the number of oracle calls required over any possible classical procedure.

\section{Towards applications}
\label{sec:dao-deep-learning}

In this last section, we sketch a potential application of finding hidden subgroups from data: distinguishing nuisance factors from task-relevant ones in data. 
This is intended to illustrate how the abstract framework developed above may find practical use.
The idea is naturally related to the framework of \textit{disentangled representation learning}, an area of machine learning research that tries to make models aware of fundamental structures in the world (see \cite{bengio2014representation, higgins2018definition, achille} and references therein).

Rather than invoke the full machinery of disentangled representations, however, we deal with the simpler problem of \emph{invariance}.
The basic idea is that, in order to distinguish task-relevant from task-irrelevant factors of variation, it is sufficient to a find a subgroup of variations that leaves performance approximately invariant.
Thus, the subgroup is ``hidden" in the data, and a quantum computer can help uncover it.

\subsection{Background}

Suppose we have a group $G_\text{prior}$ describing transformations we can apply to data. These could be active interventions (e.g. changing a parameter) or passive (e.g. conditioning on the value of a parameter).
Formally, we suppose there is a set of world states, $w \in W$, which induce probability distributions $p_w$ over a data space $\mathcal{X}$. Elements of $G_\text{prior}$ act on $W$ in the sense of a \emph{group action}, as outlined in Appendix \ref{sec:group-theory}.
We can model task-relevance using a \emph{score function} which evaluates how well a distribution $p_w$ performs at the task with a real number, $\mathcal{A}: \mathcal{X}^N \to \mathbb{R}$. Nuisance factors are transformations $g\in G_\text{prior}$ that leave the score invariant, $\mathcal{A}(w) = \mathcal{A}(g\cdot w)$.

In practice, we cannot calculate $\mathcal{A}(w)$ exactly, but rather, will compute statistics $\tilde{\mathcal{A}}_{(M)}: \mathcal{X}^M\to\mathbb{R}$ of $N$ data points sampled from $p_w$. We assume these statistics are consistent, in that $\tilde{\mathcal{A}}_{(M)} \to \mathcal{A}$ as $M \to \infty$, and choose $M$ large enough that, for any $\delta$, there is an $\epsilon = o(\delta)$ such that, with probability greater than $1-\delta$,
\[
|\tilde{\mathcal{A}}_{(M)}(\mathcal{T}_w) - \tilde{\mathcal{A}}_{(M)}(\mathcal{T}_{g\cdot w})| \leq \epsilon
\]
implies $\mathcal{A}(w) = \mathcal{A}(g\cdot w)$.
In other words, for sufficiently large data sets, if the statistics are within $\epsilon$ of each other, the scores are $\delta$-probably equal.

\subsection{Finding nuisance factors}

To find nuisance factors in practice, we can (a) sample from a distribution $p_w$ to form a training set $\mathcal{T}_w$, and (b) change distributions $w \mapsto g\cdot w$.
Part (a) is a statistical learning problem which is out of our scope. Instead we focus on (b) and suppose that $M$, $\epsilon$ and $\delta$ have been suitably chosen so that we can effectively work with the group action
\[
\mathcal{A}(w) \mapsto g \cdot \mathcal{A}(w) = \mathcal{A}(g\cdot w)
\]
of $G_\text{prior}$ on the set of real numbers $\mathcal{A}(W) \subseteq \mathbb{R}$.

\begin{example}
  Suppose the ``world'' $W$ is a dodecagon, and at each corner
  we can flip a coin (below left). The coin is $\texttt{fair}$ ($p_w(1) = 0.5$) at red corners, and $\texttt{biased}$ ($p_w(1) = 0.3$) at purple:
  \begin{center}
      \includegraphics[scale=0.27]{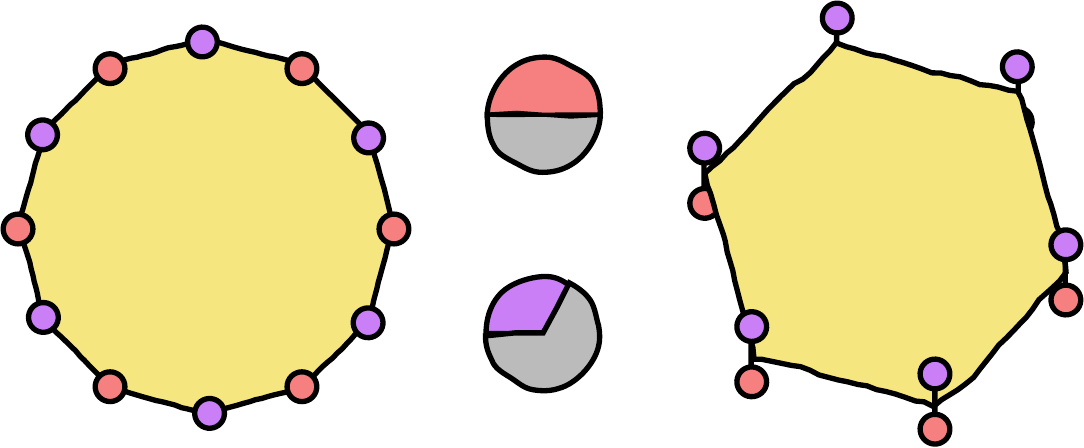}
    \end{center}
Suppose the score is given by the probability of obtaining heads $1$, or $\mathcal{A}(w)= p_w(1)$.
We can approximate this with the sample mean:
\[
\tilde{\mathcal{A}}_{(M)}(\mathcal{T}_w) = \frac{1}{M}\sum_{i=1}^M x_i.
\]
The variance of $\tilde{\mathcal{A}}_{(M)}$ is bounded by $1/4M$, and we can use to quantify our approximation of $\mathcal{A}(w)$.
\end{example}

We now focus on part (b), and how to learn the subgroup $G_\text{inv} \subseteq G_\text{prior}$ of transformations which leave the score invariant.
In the language of group actions, this is simply the \emph{stabilizer} of the current states for the action on $\mathcal{A}(W)$.
To connect this to the HSP, we simply note that the stabilizer is a hidden subgroup, with group elements $g \in G_\text{prior}$ labelled by the score $\mathcal{A}(g\cdot w)$. By making a set of random queries to $G_\text{prior}$, we can use a quantum learning algorithm---for example a heuristic built from our principle of inference to the closest invariant subspace---to guess $G_\text{inv}$. 

\begin{example}
  If we move around our dodecagonal world at random, flip coins, and label the corresponding results, we may end up with a data set
  \[
  \mathcal{T}_G = \{(0, \textcolor{red}{\bullet}), (2, \textcolor{red}{\bullet}), (3, \textcolor{violet}{\bullet})\}.
  \]
  As discussed in previous examples, this is only consistent with a hidden subgroup $G_\text{inv} = 2\mathbb{Z}_{12}$, a fact that a quantum algorithm can for example reveal by computing the DAO.
\end{example}

\section{Conclusion}
\label{sec:conclusion}

In this paper, we have approached quantum machine learning from a new and potentially fruitful angle. Instead of replacing neural networks with variational circuits on the one hand, or searching for provable quantum advantage in artificial settings on the other, we scrutinized a typical quantum computing problem to gain intuition for ways in which quantum algorithms can lead to well-motivated heuristics for learning. We identified one such mechanism, investigated a possible implementation as a cost function, and sketched an application.

These are of course only the initial steps towards a quantum algorithm for real machine learning tasks. There are many follow-up questions, such as\footnote{One might want to add the question of quantum time-complexity advantage for the learning variant of the HSP to this list. However, and as argued in the introduction, we consider this primarily of academic interest, since advantage proofs do not usually yield practical algorithms.}: 
\begin{itemize}
    \item How can we validate the idea, for example with near-term algorithms based on our proposed inference principle? 
    \item Can we use access to the annihilator for more than learning subgroups, for example to make larger machine learning models symmetry-aware?
    \item How does the principle behave if symmetries are present in ``messy'' data?
    \item Can the mechanisms analyzed here be applied to non-abelian groups, such as the symmetric group describing structures over graphs and rankings?
\end{itemize}

While these questions are deferred to future work, our investigation offers some concrete hints in the quest for pragmatic quantum advantage. First, the 
Quantum Fourier Transform seems destined to play an central role. While the QFT is crucial to many quantum algorithms, we know surprisingly little about concrete, well-compiled and possibly intermediate-term implementations for many relevant groups.

Second, the added value of quantum computers for learning might derive from their unique ability to work with symmetries via interference. Although geometric quantum machine learning involves symmetry \cite{Meyer_2023, Larocca_2022}, its predominant focus is the design of parametrized quantum circuits that are ``blind'' towards them. Our work suggests that there may be more to symmetries in quantum machine learning, especially in the realm of fault-tolerant quantum computing and targeted feature engineering. 

Thirdly, the strategy of replacing quantum oracles---which can be seen as access to full probability distributions over data---by few data samples is unexpectedly fertile.  
This way to fabricate a learning problem is arguably more faithful to the real world where we do not know, or cannot encode, the full structure of a given problem.
Access to examples of the structure, on the other hand, seems much more realistic. In this sense, moving from oracles to data---and from solving a problem to inferring the most likely solution---pushes quantum computing theory into areas that may be more relevant to applications. 

Overall, we find that machine learning, where empirical results notoriously dominate progress, challenges us to move quantum computing beyond the comfort zone of highly artificial problems to the complex demands of real computational tasks. This transition is not well explored, and requires new mathematical tools, new software capabilities and, above all, a curiosity for different research questions than the ones historically valued in quantum computing research.

\section*{Acknowledgments}
\label{sec:acknowledgments}

The authors would like to thank Joseph Bowles, Richard East, Nathan
Killoran, Korbinian
Kottman and David Wierichs for preliminary discussion and
subsequent refinement; Juan Castaneda for collaboration on
related projects; and Zhuangzi for important prior work on duality and
transformation.

\bibliography{lit}
\newpage
\appendix


\section{Group theory}
\label{sec:group-theory}

This appendix gives a brisk introduction to various definitions and
theorems from group theory we make use of in the paper and remaining
appendices.
For a deeper introduction to group theory, we recommend \cite{algebra,
  Dummit1999AbstractA, lang02}.

\subsubsection*{Basic definitions}
\label{sec:basic-definitions-1}

A \emph{group} is a set $G$ and a binary operation $\cdot: G \times G
\to G$ with the following properties:
\begin{itemize}[itemsep=0pt]
\item it is \emph{associative}, with $(g \cdot g') \cdot g'' = g \cdot (g'\cdot g'')$
  for all $g, g', g'' \in G$;
\item it has an \emph{identity} $1 \in G$ such that $g \cdot 1 = 1
  \cdot g = g$ for all $g\in G$; and
\item each element $g\in G$ has an \emph{inverse} $g^{-1}$, such that $g \cdot g^{-1} = g^{-1}\cdot g = 1$.
\end{itemize}
These laws imply that the identity and inverse are unique.
Additionally, a group is \emph{abelian} if, for
all $g, g' \in G$, $g \cdot g' = g' \cdot g$.
In this case, we denote the group operation by $+$, the identity by
$0$, and inverses by $-g$.
All groups in this paper are abelian unless stated otherwise.

A \emph{subgroup} $H \subseteq G$ is a subset which is closed under the
group operation and inverses, i.e. for all $h, h'\in H$:
\begin{equation}
  \label{eq:58}
  h + h' \in H, \quad h^{-1} \in H.
\end{equation}
It follows that $0 \in H$ and hence $H$ forms a group in its own right.
We denote subgroup inclusion by $H \leq G$.
A subgroup $H$ induces a relation on $G$ by
\begin{equation}
  \label{eq:59}
  g \sim_H g' \quad \Longleftrightarrow \quad g -g' \in H.
\end{equation}
This is an \emph{equivalence relation} since:
\begin{itemize}[itemsep=0pt]
\item (\emph{reflexivity}) $g - g = 0 \in H$;
\item (\emph{symmetry}) $h = g - g' \in H$ means
  $-h = g'-g\in H$; 
\item (\emph{transitivity}) if $h = g - g' \in H$ and $h' = g' - g''
  \in H$ then $h+h'=g - g'' \in H$.
\end{itemize}
Thus, the sets of related elements, called \emph{cosets}, are disjoint and exhaust the set.
They are all of the form
\begin{equation}
  \label{eq:60}
  r + H = \{r + h: h\in H\}
\end{equation}
for some $r$.
We let $R$ be an arbitrary set of representatives from distinct cosets.

Since the cosets are disjoint and cover $G$, the group can be written as disjoint
union of cosets, $G = \bigsqcup_{r\in R} (r + H)$.
Further, each coset has the same size, $|r + H| = |H|$, so
\begin{equation}
  \label{eq:lagrange}
  |G| = \sum_{r\in R} | r + H| = |R| |H|,
\end{equation}
and hence $|H|$ divides $|G|$. This is \emph{Lagrange's
  theorem}.

The set of cosets $\{r + H : r\in R\}$ also has the structure of a group,
with operation $(r + H) + (r' + H) = (r + r') + H$.
This is called the \emph{quotient group} and denoted $G/H$.
For general non-abelian groups this property no longer holds, and
$G/H$ is simply called the (left or right) \emph{coset space}, but we
will not worry here about the additional complications that arise.

\subsubsection*{Making and combining groups}
\label{sec:combining-groups}

A familiar abelian group is the set of \emph{integers} $\mathbb{Z}$.
The set of integer multiples $n\mathbb{Z}$ of a number $n$ is a closed under
addition and inverses, and is hence a subgroup.
The cosets are residues modulo $n$, $r + n\mathbb{Z}$. The coset
relation is more conventionally written
\begin{equation}
  \label{eq:66}
  x \in y + n\mathbb{Z} \quad \Longleftrightarrow \quad x \equiv y
  \text{ mod } n.
\end{equation}
Together, these residue cosets form the quotient group
\begin{equation}
\mathbb{Z}_n = \mathbb{Z}/n \mathbb{Z} = \{0, 1, \ldots, n-1\} = [n]\label{eq:72}
\end{equation}
of integers with addition modulo $n$.
The groups $\mathbb{Z}$ and $\mathbb{Z}_n$ have the important property
that they are \emph{cyclic}: they can be built out of single element by adding it (or its inverse) to itself some finite number
of times.
For an abelian group $G$, $g\in G$, and $k \in \mathbb{Z}$, we let
\begin{equation}
  \label{eq:67}
  k \cdot g = \overset{k \text{ times}}{\overbrace{g + g + \cdots + g}}
\end{equation}
when $k \geq 0$, and replace $g$ with $-g$ if $k < 0$.
We say a group $G$ is \emph{generated} or \emph{spanned} by a subset $\mathcal{B} \subseteq G$, and write $G =
\langle \mathcal{B}\rangle$, if every $g\in G$ can be written
\begin{equation}
  \label{eq:68}
  g = \sum_{b\in \mathcal{B}} n_b \cdot b
\end{equation}
for some integers $n_b\in\mathbb{Z}$.
This can be extended to a \emph{group presentation} $\langle
\mathcal{B}|\mathcal{I}\rangle$, where we include \emph{relations}
$\mathcal{I}$, consisting of identities satisfied by the
generators. Often, we list expressions which we take to equal the identity. Thus, we write
\begin{equation}
  \label{eq:69}
  \mathbb{Z} = \langle 1\rangle, \quad \mathbb{Z}_n = \langle 1 | n
  \cdot 1 = 0\rangle = \langle 1 | n \rangle.
\end{equation}
where the second presentation is shorthand for the first.

There are many ways to build new groups from old.
The simplest is the \emph{direct sum} of groups $G_1 \oplus G_2$, defined on the Cartesian
product $G_1 \times G_2$ by
\begin{equation}
  \label{eq:70}
  (g_1, g_2) + (g'_1, g'_2) = (g_1+g_1', g_2+g'_2).
\end{equation}
If $G_1 = \langle\mathcal{B}_1 | \mathcal{I}_2\rangle$ and $G_2 =
\langle\mathcal{B}_2|\mathcal{I}_2\rangle$, then the direct product is
generated by the disjoint union,
\begin{equation}
  \label{eq:71}
  G_1 \oplus G_2 = \langle \mathcal{B}_1 \sqcup \mathcal{B}_2 |
  \mathcal{I}_1 \sqcup \mathcal{I}_2\rangle.
\end{equation}
We can iterate this construction, so that $G^\ell$
consists of
``vectors'' $(g_1, g_2, \ldots, g_\ell)$ of $\ell$ elements $g_i \in
G$.
Although direct sum notation suggests we should
  use $\ell \cdot G$ instead $G^\ell$, we reserve this for the
  subgroup $\ell \cdot G = \{\ell \cdot g\}_{g\in G} \leq G$, where
  each element is multiplied by $\ell$, as in $\ell \mathbb{Z}_n$.

It is important to know when two groups are the same up to a
relabeling of elements.
To capture this, we use structure-preserving maps between groups.
A function $\varphi: G_1 \to G_2$ between groups is a \emph{homomorphism} if
\begin{equation}
  \label{eq:73}
  \varphi(g+_1g')=  \varphi(g)+_2 \varphi(g'),
\end{equation}
where $+_i$ is the binary operation on group $G_i$.
Such a map is an \emph{isomorphism} if its inverse (a) exists and (b)
is a homorphism.
In this case, we say the groups are \emph{isomorphic} and write $G_1
\cong G_2$.

The \emph{first
isomorphism theorem} is a helpful way to establish these. This states that, if $\varphi: G \to  G'$ is a
homomorphism, then the \emph{image} $\text{im}(\varphi) = \varphi(G)$ is isomorphic to the
quotient of $G$ by the \emph{kernel} $\text{ker}(\varphi) = \varphi^{-1}(0')$:
\begin{equation}
  \label{eq:fit}
  \text{im}(\varphi) \cong \frac{G}{\text{ker}(\varphi)}.
\end{equation}
The proof simply verifies that
$\vartheta(\varphi(g)) = g + \text{ker}(\varphi)$ is a well-defined
isomorphism $\vartheta: \text{im}(\varphi)\to G/\text{ker}(\varphi)$.
It is easy to confirm it is a homomorphism.
Note that $g, g'$ have the same image if and only if they differ by a kernel element, since $\varphi(g-g') =
\varphi(g)-\varphi(g')$ vanishes just in case $g - g' \in
\text{ker}(\varphi)$.
This means $\vartheta$ is not only well-defined, but a bijection, and
hence an isomorphism.




\subsubsection*{Independence and generators}
\label{sec:indep-gener-1}

For a group $G$, a subset $\mathcal{S} \subseteq G$ is \emph{independent} if,
for any $s\in \mathcal{S}$, the subgroup $H_s = \langle \mathcal{S}\backslash
\{s\}\rangle$ spanned by remaining elements does not include $s$.
It follows that for any subset $\mathcal{S}' \subseteq \mathcal{S}$, none of the elements
$s' \in \mathcal{S}'$ are contained in $H_{S'} = \langle \mathcal{S}\backslash \mathcal{S}'\rangle$.
This is closely related to the existence of a \emph{basis} or \emph{minimal
  generating set}, an independent set that spans $G$.
Let $\mu_\text{B}(G)$ denote the size of a maximal basis of
$G$, and $\mu_\text{I}(G)$ the
size of a maximal independent set.
It can be shown \cite{CAMERON2002641} that
\begin{equation}
  \label{eq:61}
  \mu_\text{B}(G_1 \oplus G_2) \leq \mu_\text{B}(G_1) + \mu_\text{I}(G_2),
\end{equation}
and similarly with $G_1$ and $G_2$ swapped.
Since combining a basis of $G_1$ and a basis of $G_2$ gives a basis of
$G_1 \oplus G_2$, it follows that
\begin{equation}
  \label{eq:64}
  \mu_\text{B}(G_1 \oplus G_2) \geq \mu_\text{B}(G_1) + \mu_\text{B}(G_2).
\end{equation}
If $\mu_\text{I}(G) = \mu_\text{B}(G)$ for $G_1$ and $G_2$, then we
can combine these two inequalities to get
\begin{equation}
  \label{eq:conj}
  \mu(G_1 \oplus G_2) =\mu(G_1) + \mu(G_2),
\end{equation}
where we can omit the subscript.
Unfortunately, it is not true that $\mu_\text{I}(G) =
\mu_\text{B}(G)$ in general, and there are rather elaborate non-abelian
counterexamples \cite{WHISTON2002651}.
We conjecture, however, it holds in the finite abelian case.

\subsubsection*{Group actions}
\label{sec:group-actions}

Consider a set $X$, and the set of bijections on $X$:
\begin{equation}
  \text{Perm}(X) = \{ \varphi: X \to X : \varphi^{-1} \text{
    exists}\}.\label{eq:117}
\end{equation}
This forms a (nonabelian) group under function composition.
We say that an abelian group $G$ \emph{acts on} $X$ if there is a homomorphism
$\kappa: G \to \text{Perm}(X)$, written $g \mapsto \kappa_g$.
In particular, preservation of structure means that
\[
  \kappa_{g + g'} = \kappa_g \circ \kappa_{g'}.
\]
For a direct sum of groups,
\[
  G = G_1 \oplus \cdots \oplus G_k
\]
this gives rise to tuples of functions:
\[
  \kappa_{g_1, \ldots, g_k} = (\kappa^{(1)}_{g_1}, \ldots, \kappa^{(k)}_{g_k}),
\]
where $\kappa^{(i)}$ restricts the homomorphism to factor $i$.

The point of this formalism is to let group elements act on $X$. We
will write $g \cdot x$ instead of $\kappa_g(x)$ from now on.
We first note that $0 \cdot x = x$, since for any $g \in G$,
\begin{align}
  g \cdot (0 \cdot x) & = (g + 0) \cdot x = g \cdot x.\label{eq:0-act}
\end{align}
Since each $g$ gives a bijection, we must have $0 \cdot x =x$.
The \emph{orbit} of an element $x\in X$ is its image under all the
elements of $G$:
\begin{equation}
  \label{eq:118}
  G \cdot x  = \{g \cdot x : g\in X\}.
\end{equation}
A complementary notion is the \emph{stabilizer}, which is defined as
the set of elements of $G$ that fix $x$:
\begin{equation}
  \label{eq:121}
  G^x = \{g \in G: g \cdot x = x\}.
\end{equation}
The stabilizer is always a subgroup. To see this, note it is closed under addition,
\begin{align*}
  g \cdot x = g' \cdot x & = x \quad \\
 \Longrightarrow \quad (g +
  g') \cdot x & = g \cdot (g' \cdot x) \\ & = g \cdot x \\ & = x,
\end{align*}
and taking inverses:
\begin{align*}
  g \cdot x = x \quad
 \Longrightarrow \quad (-g) \cdot x & = (-g) \cdot (g \cdot x) \\ & =
                                                                    (-g+
                                                                    g)\cdot
                                                                    x
  \\
  & = 0 \cdot x\\ & = x,
\end{align*}
where we used (\ref{eq:0-act}).
Finally, $G^x$ is nonempty since $0 \in G^x$. Thus, $G^x$ is a
subgroup as claimed.

If the stabilizer $G^x$ is a subgroup, does the orbit $G \cdot x$
have a group-intrinsic meaning? Indeed, orbits are in bijection with
\emph{cosets} of $G^x$.
More precisely, consider a coset $r + G^x$. Any element has the same image:
\[
  (r + h) \cdot x = r \cdot (h \cdot x) = r \cdot x.
\]
Conversely, if $g \cdot x = g' \cdot x$, then $g - g' \in G^x$ and
hence $g$ and $g'$ are in the same coset.
It follows that
\begin{align}
  |G\cdot x| = \frac{|G|}{|G^x|},
\end{align}
a result called the \emph{orbit-stabilizer theorem}.

In the finite abelian case, $G/G^x$ is always a subgroup with
identical cardinality, $|G \cdot x| = |G/G^x|$, and it acts
\emph{transitively} on the orbit, in the sense that we can map any two
elements $y, y' \in G \cdot x$ to each other by some $g \in G$. To see
this, simply note that
\[
  y = g \cdot x, \quad y' = g' \cdot x \quad \Longrightarrow \quad y =
  (g - g') \cdot y',
\]
so the claim is proved.

\section{Abelian characters}
\label{sec:QFT}

\subsubsection*{Basic definitions}

Consider  a finite abelian group $G$.
The \emph{dual group} $\hat{G}$ consists of all maps $\chi: G
\to \mathbb{C}$ which are \emph{multiplicative}:
\begin{equation}
  \label{eq:74}
  \chi(g + g') = \chi(g)\chi(g').
\end{equation}
It follows immediately that $\chi(0) =1$.
In a finite group, the \emph{order} $o(g) \in \mathbb{Z}_{\geq 0}$ of any element $g \in G$ is the
smallest integer such that $o(g)\cdot g = 0$.
By multiplicativity,
\begin{equation}
  \label{eq:75}
  1 = \chi(0) = \chi(o(g) \cdot g) = \chi(g)^{o(g)}.
\end{equation}
It follows that $|\chi(g)| = 1$, so every element is a phase
$e^{i\theta(g)} \in \mathrm{U}(1)$, and we
can equivalently take $\chi: G \to \mathrm{U}(1)$.

The characters in the dual group themselves form a group, hence the
name.
The operation is pointwise multiplication, with
\begin{equation}
  \label{eq:76}
  (\chi \cdot \chi') (g) = \chi(g)\chi'(g),
\end{equation}
the identity given by the constant function $\chi_0(g) = 1$,
and the inverse given by conjugation:
\begin{equation}
  \label{eq:78}
  \chi^{-1}(g) = \overline{\chi(g)}.
\end{equation}
It is easy to check that this satisfies the group axioms.

At this point, it becomes helpful to introduce the Hilbert space
$\mathcal{H}_G \cong \mathbb{C}^{G}$, where we encode the values of a
character into the coefficients of a state:
\begin{equation}
  \label{eq:85}
  |\chi\rangle = \frac{1}{\sqrt{|G|}}\sum_{g\in G} \overline{\chi(g)}|g\rangle.
\end{equation}
The \emph{shift operator} $P_s$ adds $s$ to a basis ket, $P_s|g\rangle
= |g+s\rangle$. Thus, on a character state it acts as
\begin{align}
  \label{eq:83}
  P_s |\chi\rangle & = \frac{1}{\sqrt{|G|}}\sum_{g\in G}
                     \overline{\chi(g)}|g +s\rangle \notag \\
  & = \frac{1}{\sqrt{|G|}}\sum_{g\in G}
    \overline{\chi(g-s)}|g\rangle \notag \\
    & = \frac{1}{\sqrt{|G|}}\sum_{g\in G}
                     \overline{\chi(g)} \chi(-s)^*|g\rangle = \chi(s)|\chi\rangle.
\end{align}
Thus, $|\chi\rangle$ is an eigenvector of the shift operator $P_s$
with eigenvalue $\chi(s)$.

There is a shift operator $P_s$ for each element $s\in G$, and
they all commute because the group is abelian:
\begin{equation}
  \label{eq:86}
  P_s P_{s'} = P_{s+s'} = P_{s'+s} = P_{s'}P_s.
\end{equation}
Since they all commute, they are simultaneously diagonalizable, with
an orthonormal basis of eigenvectors $|\chi\rangle$.
There are therefore $|G|$ characters in $\hat{G}$, and the states
$|\chi\rangle$ are orthogonal.

\subsubsection*{Kernel intersection}
\label{sec:kernel-intersection}

Take Fourier measurements $\hat{y}_t$, $t = 1, 2,
\ldots, T$, and define the running intersection of kernels
\begin{equation}
K = \bigcap_{t=1}^T K_{y_t}, \label{run-int}    
\end{equation}
where $K_y = \{g \in G: \chi_y(g) = 1\}$ is the subgroup annihilated
by $\chi_y$.
The Fourier measurements span an annihilator $K^\perp = \langle
\chi_{y_t}\rangle$.
If $H \neq K$, then $H > K$ and hence $K^\perp$ is smaller than $H^\perp$.
By Lagrange's theorem (\ref{eq:lagrange}), $|K^\perp|/|H^\perp| \leq
1/2$.
Thus, if the running intersection stabilizes for $c$ samples, we have
probability less than $\delta = 2^{-c}$ to sample only from $K^\perp$.
This sets the failure tolerance of our algorithm.

On the other hand, the running kernel reduces by a factor of $2$ on
every non-constant step. Combining these two observations, we learn
that, to obtain $H$ with constant success probability $1 - \delta$,
we require a number of steps $O(\log |G| \cdot \log
\delta^{-1})$, since in the worst case each running kernel except the last
stabilizes for $c - 1$ samples.

\section{Signals in Fourier space}
\label{sec:leak}

\subsubsection*{Signal-to-noise ratio}

In this appendix, we consider how incomplete training data affects Fourier sampling.
Consider a training set $\mathcal{T}$ of size $N$ for  the hidden subgroup $G$, with partial cosets $X_r \subseteq r + H$.
The probability of observing an arbitrary character $\chi \in \hat{G}$ in the standard HSP algorithm is
\begin{align}
    p_\chi = \langle\chi|\rho_\mathcal{T}|\chi\rangle
    = \frac{1}{N|G|}\sum_{r\in R}|\chi(X_r)|^2.
\end{align}
If $\chi \in H^\perp$, then $\chi(X_r)=|X_r|\chi(r)$. Hence, the probability $p_{H}$ of observing the true annihilator is
\begin{align}
    p_{H} & = \sum_{\chi\in H^\perp} p_\chi\notag\\
    & = \sum_{\chi\in H^\perp, r\in R}\frac{|X_r|^2}{N|G|} |\chi(r)|^2 \notag \\
    & = \frac{|H^\perp|}{N|G|}\sum_{r\in R}|X_r|^2\notag\\
    & = \frac{\Vert\mathbf{X} \Vert_2^2}{N|H|}, \label{eq:p_H}
\end{align}
where $\mathbf{X} = (|X_r|)_{r\in R}$ is the vector of partial coset sizes. If we think of observing the annihilator as signal, and characters outside the annihilator as noise, the signal-to-noise ratio is
\begin{align}
    \text{SNR}(\mathcal{T}) & = \frac{p_{H}}{1-p_{H}} = \frac{\Vert\mathbf{X} \Vert_2^2}{N|H| - \Vert\mathbf{X} \Vert_2^2}.
\end{align}
Norm inequalities show that
\[
\frac{N^2|H|}{|G|}\leq \Vert \mathbf{X}\Vert_2^2 \leq N^2,
\]
so that $\Vert \mathbf{X}\Vert_2^2 = \Theta(N^2)$. If $N$ is much smaller than $|H|$, we have $\text{SNR}(\mathcal{T}) = \Theta(N)$, so the ratio is linear in the sample size as claimed.

\subsubsection*{False signals}

Instead of noise, we can view characters outside $H^\perp$ as giving \emph{false signals}. They suggest falsely that they are contained in the annihilator of the hidden subgroup. The basic observation is that, if $\chi \notin H^\perp$, then
\[
\frac{1}{|H|}\chi(r + H) = 0.
\]
We can view this as the statement that the random variable $\chi(g)$ for $g \in r + H$ chosen uniformly at random has average zero. The variance is unity:
\[
\frac{1}{|H|}\sum_{h\in H}|\chi(r+h)|^2 = 1.
\]
If we view a partial coset $X_r$ as chosen at random without replacement from $r+H$, then to a reasonable approximation $\chi(X_r)$ has vanishing mean and variance $|X_r|$.

For a subgroup $\tilde{H} \neq H$, modifies (\ref{eq:p_H}) as follows:
\begin{align}
    p_{\tilde{H}} & = \sum_{\tilde{\chi}\in\tilde{H}^\perp, r\in R} \frac{1}{N|G|} |\chi(X_r)|^2 \notag \\
    & = \frac{1}{N|G|}\sum_{r\in R}\left[\sum_{\tilde{\chi}\in\tilde{H}^\perp_0} |\chi(X_r)|^2 + \sum_{\tilde{\chi}\in\tilde{H}^\perp_\cap}|X_r|^2\right] \notag \\
    & \approx \frac{|\tilde{H}^\perp_0|\Vert \mathbf{X}\Vert_1 }{N|G|} + \frac{\Vert \mathbf{X}\Vert_2^2}{N|\tilde{H}_\cap|} \notag \\
    & = \frac{1}{|\tilde{H}|} - \frac{1}{|\tilde{H}_\cap|} + \frac{\Vert \mathbf{X}\Vert_2^2}{N|\tilde{H}_\cap|}
\end{align}
since $\Vert \mathbf{X}\Vert_1 = N$.
This is larger than the probability of the true signal $p_H$ just in case
\[
\frac{1}{|\tilde{H}|} - \frac{1}{|\tilde{H}_\cap|} \geq \frac{\Vert \mathbf{X}\Vert_2^2}{N}\left(\frac{1}{|H|} - \frac{1}{|\tilde{H}_\cap|}\right).
\]
Since $\Vert \mathbf{X}\Vert_2^2 = \Theta(N^2)$, this would require
\[
\frac{\tilde{a}-1}{a - 1} = \Omega(N),
\]
where $\tilde{a}=|\tilde{H}_\cap|/|\tilde{H}|$ and $a=|\tilde{H}_\cap|/|H|$ are integers by Lagrange's theorem.
For large $a, \tilde{a}$, this holds only for smaller and smaller guesses, $|\tilde{H}|= \Omega(|H|/N)$, and becomes increasingly unlikely as our training set increases.

\section{PAC learning}
\label{sec:sample-eq}

\subsubsection*{Varieties of oracle}
\label{sec:varieties-oracle}

Classical PAC learning uses a classical example oracle $\mathsf{EX}(c,
\mathcal{D}) = (x, f_c(x)) \in \mathcal{X}\times \mathcal{Y}$ satisfying
\begin{equation}
  \mathbb{P}[\mathsf{EX}(c, \mathcal{D}) = (x, f_c(x))] = \mathcal{D}(x).
\end{equation}
Quantum PAC learning uses a quantum example oracle:
\begin{equation}
  \label{eq:qex2}
  \big|\mathsf{QEX}(c, {\mathcal{D}})\big\rangle = \sum_{x\in
    \mathcal{X}}\sqrt{\mathcal{D}(x)}|x, f_c(x)\rangle.
\end{equation}
Measuring the quantum example in the computational basis yields a classical
example $\mathsf{EX}(c, \mathcal{D})$.

We can select $N$ random inputs $x\in \mathcal{X}$ with a
distribution $\mathcal{D}$, with a vector $\mathbf{n} = (n_x)_{x\in
  \mathcal{X}}$ of counts which add to $N$ and are multinomially distributed:
  \begin{equation}
\mathbb{P}[\mathbf{n}] =
  \binom{N}{\mathbf{n}}\prod_{x\in\mathcal{X}} \mathcal{D}(x)^{n_x}.\label{eq:88}
\end{equation}
A \emph{quantum training state} can either fix $\mathbf{n}$,
\begin{align}
  \label{eq:QTS}
  \big|\mathsf{QTS}(c, {\mathcal{D}}, \mathbf{n})\big\rangle = \sum_{x
  \in\mathcal{X}} \sqrt{\frac{n_x}{N}} |x, f_c(x)\rangle.
\end{align}
or fix $N$ and choose $\mathbf{n}$ randomly on each call, which we
refer to as $\big|\mathsf{QTS}_N(c, {\mathcal{D}})\big\rangle$.
Both quantum training states approach the quantum example
$|\textsf{QEX}\rangle$ as $N \to \infty$, since
\begin{equation}
  \label{eq:lln}
  \lim_{N\to \infty} \frac{n_x}{N} = \mathcal{D}(x),
\end{equation}
almost surely, by the law of large numbers.


The training state (\ref{eq:psi_T}) is an even simpler
type of training oracle we called a \emph{uniform training state}:
\begin{align}
  \label{eq:UTS}
  \big|\mathsf{UTS}(c, {\mathcal{D}}, \mathbf{b})\big\rangle = \sum_{x
  \in\mathcal{X}} \frac{b_x}{\sqrt{N}} |x, f_c(x)\rangle,
\end{align}
for a one-hot vector $\mathbf{b}=(b_x)_{x\in\mathcal{X}} \in \{0,
1\}^{\mathcal{X}}$ of weight $N$, $\Vert \mathbf{b}\Vert_1 = N$, where
we sample elements without replacement. Similarly, we can replace a
fixed vector with $N$ random calls, $\big|\mathsf{UTS}_N(c,
{\mathcal{D}})\big\rangle$.

Since we sample without replacement, $N \leq |\mathcal{X}|$. For $N =
|\mathcal{X}|$, assuming each item has nonzero probability, we do not
recover the quantum example but rather the
\emph{uniform example}:
\begin{align}
  \big|\mathsf{UTS}_{|\mathcal{X}|}(c)\big\rangle & =
  \big|\mathsf{UEX}(c, {\mathcal{D}})\big\rangle \notag \\ & =
  \frac{1}{\sqrt{|\mathcal{X}|}}\sum_{x\in\mathcal{X}} |x, f_c(x)\rangle.   \label{eq:94}
\end{align}
This is the resource of interest in exact quantum learning
\cite{arunachalam2017survey}, since we obtain the uniform example by
applying an exact membership oracle to a uniform superposition of
inputs.
In this sense, the uniform training state is the natural
``deformation'' to consider in going from the exact HSP problem to a
learning variant.

The advantage of the PAC formulation is that it accomodates a wider
range of distributions. We split the difference, using the PAC
formulation to discuss sample complexity and uniform training
for the learning algorithm itself. To ensure these two
approaches mesh, we make strong enough distributional
assumptions that we can learn the hidden subgroup using states of the
form (\ref{eq:UTS}), with enough flexibility to saturate the sample
complexity bounds.

\subsubsection*{VC dimension and independent sets}

In this appendix, we compute the VC dimension of a finite abelian group in the form
(\ref{eq:ftfag}):
\[
  G \cong \mathbb{Z}_{q_1}^{\ell_1} \oplus \mathbb{Z}_{q_2}^{\ell_2}
  \oplus \cdots \oplus \mathbb{Z}_{q_M}^{\ell_M},
\]
where $q_i = p_i^{m_i}$.
Let  us reformulate VC dimension in terms of independent sets in a group.
For a subset $\Gamma \subseteq G \times G$, we can form the difference
set
\begin{equation}
  \label{eq:51}
  \Gamma_\text{diff} = \{g'-g: (g, g') \in \Gamma\}.
\end{equation}
The concept class $\mathsf{C}_G$ shatters $\Gamma$ just in case
$\Gamma_\text{diff} \subseteq G$ is independent.
To see this, note that for $\Gamma' \subseteq
\Gamma_\text{diff}$,
\begin{equation}
  \label{eq:56}
  H_{\Gamma'} \cap \Gamma_\text{diff} = \Gamma_\text{diff}\backslash \Gamma'.
\end{equation}
It follows that the subgroups shatter $\Gamma$, as required.
Conversely, any shattered set $\Gamma$ must have this property.
Thus, $\text{dim}_\text{VC}(\mathsf{C}_G) = \mu_\text{I}(G)$.

Our earlier conjecture (\ref{eq:conj}) for finite abelian groups
implies that the VC dimension is additive for direct sums:
\begin{align}
\text{dim}_\text{VC}(\mathsf{C}_{G_1 \oplus G_2}) & = \mu_\text{I}(G_1
                                                    \oplus G_2) \notag
  \\
  & = \mu_\text{I}(G_1) +\mu_\text{I}(G_2) \notag \\
& = \text{dim}_\text{VC}(\mathsf{C}_{G_1}) + \text{dim}_\text{VC}(\mathsf{C}_{G_2}).\label{eq:17}
\end{align}
For a single cyclic factor $V = \mathbb{Z}_q$ for $q = p^m$, a basis
is clearly given by $\mathcal{B} = \{1\}$. By the Burnside basis
theorem \cite{MCDOUGALLBAGNALL2011332}, every basis and independent
set is the same size.
Thus, $\mu_\text{B}(V) = \mu_\text{I}(V) = 1$.
Together, these results imply
\begin{equation}
  \label{eq:20}
  \text{dim}_\text{VC}(G) = \sum_{i=1}^M \ell_i,
\end{equation}
which we used for sample complexity in \S\ref{sec:sample}.

\section{Cost function details}
\label{sec:cost-funct-deta}

Here, we collect basic results for evaluating the DAO cost function.

\subsubsection*{Complete data}
\label{sec:choice-dcota-regul}

For full data $X = G$, the partial cosets become the original cosets, $X_r = r + H$ (which corresponds to access to the full oracle).
Then the DAO length term in (\ref{eq:dao-reg}) becomes
\begin{align}
  \Vert \beta(\tilde{H})\Vert_2^2  & = \frac{|H|}{|G|}\sum_{r\in R}\left|\langle r+H|\tilde{H}^\perp\rangle\right|^2 \notag \\ & = \frac{1}{|G|^2}\sum_{r\in R}\left|\sum_{\hat{y}\in
    \tilde{H}^\perp}\chi_y(r + H)\right|^2 \notag \\
  & = \frac{1}{|R|^2}\sum_{r\in R}\left|\sum_{\hat{y}\in
    \tilde{H}^\perp\cap H^\perp}\chi_y(r)\right|^2.
\end{align} 
At this point, we can exploit symmetry of the Fourier transform, $\chi_y(r) = \chi_r(y)$, which follows from using the usual characters $\chi_x(y)\propto e^{-2\pi i xy/n}$ on cyclic components and multiplying characters and using (\ref{eq:ftfag}). Then, defining $\tilde{H}_\cap^\perp = \tilde{H}^\perp \cap H^\perp$ and $\tilde{H}_\cap = (\tilde{H}_\cap^\perp)^\perp$, we have
\begin{align}
  \Vert \beta(\tilde{H})\Vert_2^2
  & = \frac{1}{|R|^2}\sum_{r\in R}\left|\sum_{\hat{y}\in
    \tilde{H}^\perp\cap H^\perp}\chi_r(y)\right|^2 \notag \\
    & = \frac{1}{|R|^2}\sum_{r\in R}\left|\chi_r(\tilde{H}_\cap^\perp)\right|^2 \notag \\
    & = \frac{|\tilde{H}_\cap^\perp|^2 |\hat{R} \cap \tilde{H}_\cap|}{|R|^2},
\end{align} 
as claimed in (\ref{eq:full-1}).

\subsubsection*{Sparse data}

We can use similar techniques in the case of sparse data. First, we perform the decomposition
\begin{align}
    \Vert \beta(\tilde{H})\Vert_2^2 & = \sum_{r \in R}\frac{1}{N|G||X_r|}\left|\sum_{\hat{y}\in \tilde{H}^\perp} \chi_y(X_r)\right|^2 \notag \\
    & = \sum_{r \in R}\frac{1}{N|G||X_r|}\left||X_r| \cdot\chi_r(\tilde{H}^\perp_\cap) + \sum_{\hat{z}\in \tilde{H}^\perp_0} \chi_z(X_r)\right|^2 \label{eq:partial-dao-0},
\end{align}
where the characters $\tilde{H}^\perp_\cap = \tilde{H}^\perp\cap H^\perp$ experience positive interference and the remainder $\tilde{H}^\perp_0 = \tilde{H}^\perp \backslash \tilde{H}^\perp_\cap$ tend to cancel out.
As usual, the constructive term gives
\[
|X_r| \cdot\chi_r(\tilde{H}^\perp_\cap) = |X_r||\tilde{H}^\perp_\cap| \cdot \mathbb{I}[r \in \tilde{H}_\cap].
\]
To evaluate the second part, we will make the heuristic assumption that $\chi_z(X_r)$ acts as a random sample of the phases $\chi_z(g)$, with vanishing mean (since $\chi_z(H) = 0$) and variance $|X_r|$. Moreover, we assume that each $\chi_z$ is an independent random variable.

All of these assumptions depend on the sampling process, and will be subject to more careful analytic and numerical checks in subsequent work. For now, we simply identify
\[
\sum_{\hat{z}\in \tilde{H}^\perp_0} \chi_z(X_r) = \sqrt{|X_r||\tilde{H}^\perp_0|}\epsilon_0(r),
\]
where we treat $\epsilon_0$ as a complex random variable of zero mean and unit variance.
Substituting these into (\ref{eq:partial-dao-0}) and expanding the square, we get three terms:
\begin{align}
    \Vert \beta(\tilde{H})\Vert_2^2 & \approx \alpha_1 + \alpha_2 + \alpha_3.
\end{align}
The first is the purely constructive term.
Assuming the partial cosets are roughly equal in size, we can replace $|X_r| \approx N/|R|$ and use (\ref{eq:rand-sampler}) to get:
\begin{align}
\alpha_1 & = \sum_{r\in R}\frac{|\tilde{H}_\cap^\perp|^2 |X_r|}{N|G|} \mathbb{I}[\hat{r}\in \tilde{H}_\cap] \notag \\
    & \approx \frac{|\tilde{H}_\cap^\perp|^2 |\hat{R}\cap \tilde{H}_\cap|}{|R||G|}  \notag \\
    & \approx \frac{|\tilde{H}_\cap^\perp|}{|G|}.
\end{align}
As before, this term is maximized by maximizing the overlap with $H^\perp$.

We then have a cross-term, $\alpha_2$. Assume again that $|X_r| \approx N/|R|$, (\ref{eq:rand-sampler}), and that the error term $\epsilon_0(r)$ is roughly constant over $r$. Then
\begin{align*}
    \alpha_2 & = \sum_{r\in R} \frac{|\tilde{H}_\cap^\perp| \sqrt{|X_r||\tilde{H}_0^\perp|}}{N|G|} \cdot \mathbb{I}[\hat{r}\in \tilde{H}_\cap]\cdot 2\Re[\epsilon_0(r)] \notag \\
    & \approx \frac{|\tilde{H}^\perp_\cap|}{|G|}\sqrt{\frac{|\tilde{H}_0^\perp| |R|}{N}} \cdot 2\Re[\bar{\epsilon}_0], \label{eq:alpha_2}
\end{align*}
where $\bar{\epsilon}_0 = \sum_r\epsilon_0(r)/|R|$.
Finally, we have the squared error term, $\alpha_3$, with
\begin{align}
    \alpha_3 & = \sum_{r\in R}\frac{|X_r||\tilde{H}^\perp_0| |\epsilon_0(r)|^2}{N|G||X_r|} \notag \\
    & \approx \frac{|\tilde{H}^\perp_0|(|R|-1)}{N|G|}\cdot \overline{\mbox{var}}_r[\epsilon_0],
\end{align}
where we write the sum over $r$ as a sample variance $\overline{\mbox{var}}$.

We see that terms involving the random error are multiplied by powers by $|R|/N$. This means that, if $|R| > N$, error terms can potentially dominate the cost function, but if $|R| < N$ they will tend to be suppressed. This suggests we need $N = O(|R|)$ to have the constructive interference term dominate.
The second correction can have either sign, but the third correction is positive and favours larger $\tilde{H}_0^\perp|$ and hence $|\tilde{H}^\perp|$. We need to choose a large enough regularization to counteract this.

\vfill

\end{document}